\documentclass[10pt,a4paper]{iopart}
\usepackage{textcomp}
\usepackage{lmodern}     
\usepackage{iopams}
\expandafter\let\csname equation*\endcsname\relax
\expandafter\let\csname endequation*\endcsname\relax
\usepackage{amsmath}
\usepackage{amssymb}
\usepackage{amsthm}
\usepackage{amsfonts}
\usepackage{graphicx}
\usepackage{color}
\usepackage{bbm}
\usepackage{float}
\usepackage{dsfont}
\usepackage{feynmp}
\usepackage{tikz}
\usepackage{drawInt}
\usepackage[breaklinks=true,colorlinks=true,linkcolor=blue,urlcolor=blue,citecolor=blue]{hyperref}
\usepackage{subcaption}

\newcommand{\id}[1] {\ensuremath{\text{d} #1 \;}}           
\renewcommand{\d}[3]{\ensuremath{\frac{\text{d}^{#1} #2}{\text{d} #3^{#1}}}} 
\newcommand{\pd}[3]{\ensuremath{\frac{\partial^{#1} #2}{\partial #3^{#1}}}}  
\newcommand{\at}[1]{\ensuremath{\big|_{#1} } }

\newcommand{\ket}[1]{\ensuremath{\left| #1 \right>}}
\newcommand{\bra}[1]{\ensuremath{\left< #1 \right|}}
\newcommand{\braket}[2]{\ensuremath{\left< #1 \ \vphantom{#2} \right| 
\left. #2 \vphantom{#1} \right>}}

\newcommand{\matrixel}[3]{\ensuremath{\left< #1 \vphantom{#3} \right| #2 
\left| #3 \vphantom{#1} \right>}}

\renewcommand{\exp}[1]{\ensuremath{ \; \text{exp} \left( #1 \right) } }

\newcommand{\abs}[1]{\left| #1 \right|}

\renewcommand{\det}[1]{\ensuremath{\text{det}\left( #1 \right) }}

\newcommand{\eps}{\ensuremath{\varepsilon}}
\newcommand{\bral}{\ensuremath{\bra{\{\lambda\}}}}

\newcommand{\ketl}{\ensuremath{\ket{\{\lambda\}}}}
\newcommand{\ketm}{\ensuremath{\ket{\{\mu\}}}}
\newcommand{\rom}[1]{\uppercase\expandafter{\romannumeral #1\relax}}
\def\nn{\nonumber\\}
\def\fr#1{(\ref{#1})}
\newcommand{\bea}{\begin{eqnarray}}
\newcommand{\eea}{\end{eqnarray}}
\newcommand{\be}{\begin{equation}}
\newcommand{\ee}{\end{equation}}

\begin{document}
\title[Spinon Decay]{Spinon decay in the spin-1/2 Heisenberg chain
with weak next nearest neighbour exchange} 
\author{Stefan Groha}
\address{The Rudolf Peierls Centre for Theoretical Physics, University of Oxford, Oxford, OX1 3NP, United Kingdom}
\author{Fabian H.L. Essler}
\address{The Rudolf Peierls Centre for Theoretical Physics, University of Oxford, Oxford, OX1 3NP, United Kingdom}

\begin{abstract}
Integrable models support elementary excitations with infinite
lifetimes. In the spin-1/2 Heisenberg chain these are known as
spinons. We consider the stability of spinons when a weak
integrability breaking perturbation is added to the Heisenberg
chain in a magnetic field. We focus on the case where the perturbation
is a next nearest neighbour exchange interaction. We calculate the
spinon decay rate in leading order in perturbation theory using
methods of integrability and identify the dominant decay channels. The
decay rate is found to be small, which indicates that spinons remain
well-defined excitations even though integrability is broken.
\end{abstract}


\maketitle
\section{Introduction}
Integrable many-particle quantum systems are special in that they
support stable elementary excitations. These typically are related in
very complicated ways to the basic degrees of freedom. For example, in
the Heisenberg antiferromagnet the elementary excitations are
interacting spin-1/2 objects called spinons \cite{Faddeevspinon,FT}. 
Crucially, these elementary excitations are protected from decay into
multi-particle excitations by the existence of local integrals of
motion, even in cases where decay is kinematically allowed. In such
situations application of integrability breaking perturbations has the
immediate effect of inducing particle decay, and an important question
is how large the corresponding decay rates are. If they are small, the
elementary excitations of the integrable model will remain a good
basis for describing the physics of the perturbed model. Such
questions have been investigated in some detail for integrable quantum
field theories \cite{TCSA,DM,DGM,PT}. The case of integrable lattice
models is considerably harder, and to the best of our knowledge has
not been investigated so far. The added difficulty compared to
field theory cases is that the description of the ground and excited
states is more complicated (see below). The question of what effects
weak integrability breaking perturbations have on the excitation
spectrum of lattice models is also of importance in so-called mobile
impurity approaches to the calculation of threshold singularity
exponents in lattice models \cite{ISG}. As pointed out in
Ref.~\cite{EPS} in the context of the Hubbard model, there exist
different formulations of mobile impurity models \cite{SIG1}, which
correspond to different choices of bases of elementary excitations.
One may argue that for integrable models the ``integrable'' basis of
elementary excitations ought to be the preferred choice. An obvious
question is then whether this remains the case even if integrability
is weakly broken. This is intimately related to how large the decay
rate of the excitations is once a perturbation is applied. For the
Hubbard model the available integrable model technology \cite{book}
does not currently permit to answer this question. In this work we
therefore consider the simpler case of the spin-$1/2$ Heisenberg XXZ
chain of length $L$ in a magnetic field $h$
\be
\fl\qquad
H(J,\Delta,h)=J\sum_{j=1}^L \left(S_j^x S_{j+1}^x +S_j^y S_{j+1}^y
+\Delta S_j^z S_{j+1}^z\right) - h \sum_{j=1}^L S_j^z\ .
\label{XXZhamiltonian} 
\ee
Here $J<0$ and $S_j^\alpha$ are spin operators with commutation relations
\begin{align}
[S_i^\alpha,S_j^\beta]=i\delta_{ij}\eps_{\alpha\beta\gamma}S_i^\gamma\ .
\label{spinalgebra}
\end{align}
The spectrum of \fr{XXZhamiltonian} is gapless for $\abs{\Delta}\leq
1$ and $|h|<J(1+\Delta)$ \cite{takahashi,korepin}. The model is
integrable and elementary excitations over the ground state carry
$S^z$ quantum number $1/2$ and are known as "spinons"
\cite{Faddeevspinon,FT,takahashi,korepin}. A simple way of perturbing the 
model away from the integrable point is by introducing a next nearest
neighbour interaction
\begin{align}
\delta H= \kappa \sum_j S_j^z S_{j+2}^z\ .
\end{align} 
This interaction destroys integrability, but still commutes with the
total spin operator along the z-axis $S^z=\sum_j S_j^z$. Hence the
z-component of the total spin remains a good quantum number. In
presence of the perturbation spinons cease to be exact elementary excitations
and we expect them to acquire a finite life-time. Using Fermi's golden
rule for small perturbations the decay rate can be expressed in the form
\begin{align}
  \Gamma=2\pi \sum_{f} \abs{\mathcal{M}(i\to f)}^2 \rho_f(E_i) \;\delta_{p_f,p_i} \label{Fermigoldenrule}
\end{align}
where $E_f$ and $p_f$ ($E_i$ and $p_i$) are the energy and momentum of
the final (initial) state \cite{peskin}, $\rho_f(E)$ the density of
states of the final state and the matrix element $\mathcal{M}$
is given by
\begin{align}
\mathcal{M}=\bra{f}\delta H \ket{i}\ .
\end{align}
We are interested in the case where the initial state is an exact one-spinon
eigenstate of \eqref{XXZhamiltonian}, while the final state is any
exact eigenstate of the unperturbed system. 

Most of our analysis will focus on the isotropic Heisenberg model at
$\Delta=1$ and $h>0$. Other values of $\Delta$ can be treated in the same way
and $h<0$ by considering the spin overturned sector. 
The outline of this paper is as follows. Section \ref{sec:Bethe} presents
a brief summary of the Bethe Ansatz solution of the Heisenberg
model. We then describe the excited states that contribute to the
decay rate in section \ref{sec:excitations}.  We then use the
Algebraic Bethe Ansatz to obtain explicit expressions
for the matrix elements describing the spinon decay, \emph{cf.}
\ref{sec:matrixel}. In section \ref{sec:decay} we then numerically
determine the contributions of various decay channels to the decay
rate. We end with a discussion of our results in section \ref{sec:summary}.

\section{Bethe Ansatz solution of the XXX-chain} 
\label{sec:Bethe}
\subsection{Coordinate Bethe Ansatz}
Eigenstates of the XXZ Hamiltonian \eqref{XXZhamiltonian} can be constructed
by means of the Bethe Ansatz \cite{BetheXXX} for any value of the anisotropy
$\Delta$. As $S^z$ commutes with the XXZ Hamiltonian and with the
perturbation $\delta H$, it is convenient to work in a sector with a fixed
number of down-spins $N$ with respect to the ferromagnetic state  
\begin{align}
\ket{0}:=\ket{S^z=\frac{L}{2}}=\bigotimes_{j=1}^L \ket{\uparrow}_j\ ,
\end{align}
which will be used as a reference state in the following. 
Energy eigenstates with $N$ down-spins take the form
\begin{align}
\ket{N}=\sum_{j_1,\dots,j_N=1}^L a(j_1,\dots,j_N) \prod_{a=1}^N
S_{j_a}^{-} \ket{0}\ ,
\label{generalstateform} 
\end{align}
where $1\leq j_1 < j_2 < \dots < j_N \leq L$. The wave functions have
Bethe Ansatz form \cite{takahashi,korepin,Gaudin,Orbach1} 
\begin{align}
a(j_1,\dots,j_N)=&\sum_{P\in S_N} (-1)^{[P]} \mathcal{A}_P
\exp{i\sum_{a=1}^N k_{Pa}j_a}\ ,\nn
\mathcal{A}_P =&\prod_{a<b}\left(e^{i(k_{P_a}+k_{P_b})}+1-2\Delta e^{ik_{P_a}}\right).
\label{bethewavefunction} 
\end{align}
The energy of the state with wave function \fr{bethewavefunction} is
given by
\begin{align}
  E&=J\sum_{a=1}^N (\cos{k_a}-\Delta)-h\left(\frac{L}{2}-N\right). \label{Energ}
\end{align}
\subsection{Bethe equation for the XXX model}
Imposing periodic boundary conditions on the wave functions
\eqref{bethewavefunction} leads to quantization conditions for the
wave numbers $k_a$ known as \emph{Bethe Ansatz equations}
\begin{align}
e^{ik_a L}=\prod_{a\neq b}^{N}\left[ -\frac{2\Delta e^{ik_a}-e^{ik_a+ik_b}-1}{2\Delta e^{ik_b}-e^{ik_a+ik_b}-1} \right].
\end{align}
From here on we set $\Delta=1$. It is convenient to introduce
rapidity variables $\lambda_a$ defined by
\begin{align}
e^{ik_a}=\frac{\lambda_a-i/2}{\lambda_a+i/2}.
\end{align}
In terms of the rapidity variables the Bethe Ansatz equations read
\begin{align} 
\left(\frac{\lambda_a-i/2}{\lambda_a+i/2}\right)^L=\prod_{\substack{b=1\\ b\neq
    a}}^N \frac{\lambda_a-\lambda_b-i}{\lambda_a-\lambda_b+i}\ ,\
a=1,\dots,N.
\label{Bethesinh}
\end{align}
A standard way of analyzing \fr{Bethesinh} is by employing the
\emph{string hypothesis}. This assumes that all solutions of
(\ref{Bethesinh}) are composed of strings of the form
\be
\lambda^{n,j}_\alpha=
\lambda^{n}_\alpha+\frac{i}{2}(n+1-2j)+\delta^{n,j}_\alpha\ ,\quad
j=1,\dots, n.
\label{strings}
\ee
Here $\delta^{n,j}_\alpha$ are deviations from ``ideal'' strings and
are assumed to be exponentially small in system size. 
Let us now consider a solution to \fr{Bethesinh} that contains $M_n$
strings of length $n$ with corresponding string centres
$\lambda^{n}_\alpha$ (this implies that $\sum_nM_nn=N$). 
Substituting \fr{strings} into \fr{Bethesinh} and
neglecting the deviations we obtain a set of coupled equations for the 
set $\{\lambda^n_\alpha\}$. Taking logarithms we arrive at
\be
L\theta\left(\frac{\lambda^n_\alpha}{n}\right)=2\pi
  I^n_\alpha
+\sum_{(m,\beta)\neq(n,\alpha)}\theta_{nm}(\lambda^n_\alpha-\lambda^m_\beta).
\label{LBAE}
\ee
Here $I^n_\alpha$ are integer or half-odd integers numbers (arising
from taking logarithms), $\theta(x)=2{\rm arctan}(2x)$, and 
\be
\theta_{nm}(x)=\begin{cases}
\theta\big(\frac{x}{2n}\big)
+2\sum_{j=1}^{n-1}\theta\big(\frac{x}{2j}\big) & \text{for } m=n\\
\theta\big(\frac{x}{|n-m|}\big)
+2\theta\big(\frac{x}{|n-m|+2}\big)+\ldots
+2\theta\big(\frac{x}{n+m-2}\big)+
\theta\big(\frac{x}{n+m}\big) & \text{for } m\neq n
\end{cases}.
\ee
Equations \fr{LBAE} are called \emph{Takahashi's equations}. They
relate the solutions of the BAE to a set of integer of half-odd
integer numbers, which therefore can be considered as quantum numbers
of our problem. The permitted ranges of the $I^n_\alpha$
are \cite{takahashi} 
\be
|I^n_\alpha|\leq\frac{1}{2}\left[L-1-\sum_{m=1}^\infty \left(
2{\rm min}(m,n)-\delta_{n,m}\right)M_m\right].
\ee
Energy and momentum of solutions to \fr{LBAE} are given by
\bea
E&=&\sum_{m=1}^\infty\sum_{\beta=1}^{M_m} \left(-\pi
Ja_n(\lambda^m_\beta)+mh\right)-\frac{hL}{2}\ ,\nn
P&=&\pi+\sum_{m=1}^\infty\sum_{\beta=1}^{M_m} \frac{2\pi I^m_\beta}{L}
\label{EP}
\eea
where we have defined
\be
a_n(x) =\frac{1}{2\pi}\frac{n}{x^2+(n/2)^2}\ .
\ee
All solutions to Takahashi's equations correspond to highest weight states
of the spin SU(2) algebra \cite{FT}
\be
S^+|\{\lambda^{n}_\alpha\}\rangle=0.
\ee
A complete set of energy eigenstates is then obtained by acting with
the spin lowering operator on these highest weight states
\be
\left(S^{-}\right)^m|\{\lambda^{n}_\alpha\}\rangle\ ,\quad
m=0,1,\dots, L-2\sum_{n}nM_n.
\label{descendant}
\ee
\section{Low lying excitations and spectrum}
\label{sec:excitations}
In order to have access to single-spinon excitations we need to
consider odd chain lengths $L$. For even values of $L$ the lowest
excitations involve at least two spinons \cite{FT}. 
\subsection{One particle and one hole excitations}
For odd $L$ with an odd number $N$ of down spins there are two
degenerate lowest energy states. They are obtained by considering real 
solutions (1-strings) to the Bethe Ansatz equations and choosing either
\be
I^1_\alpha=-\frac{N}{2}+\alpha\ ,\quad\alpha=1,\dots,N,
\ee
or
\be
I^1_\alpha=-\frac{N}{2}+1+\alpha\ ,\quad\alpha=1,\dots,N,
\ee
The corresponding configurations of half-odd integers for $N=11$ look
as follows:
\begin{center}
        \halfintegers
        \border
        \intrange{-9}{10}
        \fillI{-4}{5}
        \particleadd{6}{}
        \intLabel{I_\alpha^1}
        \render
\end{center}
\begin{center}
        \halfintegers
        \border
        \intrange{-9}{10}
        \fillI{-4}{5}
        \particleadd{-5}{}
        \intLabel{I_\alpha^1}
        \render
\end{center}

The energy density $e(h)$ of these two states in the
thermodynamic limit 
\be
L,N\to \infty, \frac{N}{L}=n=\text{fixed}
\ee
can be expressed in terms of the solution of a
linear integral equation for the root density $\rho_1(\lambda)$, 
\emph{cf.} Ref.~\cite{takahashi}
\be 
\rho_1(\lambda)=
a_1(\lambda)-\int_{-B}^B \id \eta a_2\left(\lambda-\eta\right)\
\rho_1(\eta).
\ee
Here the integration boundary $B$ is determined by the density of down
spins $n$ through
\be
\int_{-B}^B \id\lambda \rho_1(\lambda) = n.
\ee
The energy per site is then given by
\be
e(h) = \int_{-B}^B \id\lambda \rho_1(\lambda) \eps^{(0)}_1(\lambda) \ ,
\ee
where 
\be
\eps^{(0)}_1(\lambda)=-J\pi a_1(\lambda)+h\ .
\ee

The two states above are particular limits of one-parameter
particle-like and hole-like excitations. The particle excitation 
corresponds to $I^1_\alpha$ configurations of the form
\begin{center}
        \noint
        \border
        \intrange{-9}{10}
        \fillI{-4}{5}
        \particleadd{8}{I^{p}}
        \intLabel{I_\alpha^1}
        \render
\end{center}
whereas the hole-like excitation is obtained by promoting one half-odd
integer $I_\alpha^1$ in the ground state configuration to the ``Fermi edge''
that has one fewer half-odd integer:
\begin{center}
        \noint
        \border
        \intrange{-9}{10}
        \fillI{-4}{5}
        \holeadd{1}{I^{h}}
        \particleadd{-5}{}
        \particleadd{6}{}
        \intLabel{I_\alpha^1}
        \render
\end{center}
Both types of excitations involve a single parameter: $I^p$ for the
particle excitation and $I^h$ for the hole excitation. For
asymptotically large system sizes $L$ the energies and momenta of
these excitations are given by \cite{takahashi}
\begin{align}
  E^p &= Le + \eps_1(\lambda^p) +o(1), \qquad \abs{\lambda^p}>B\ , 
  \label{Eq:particleen}\\
P^p &= \pi + 2\pi \int_0^{\lambda^p} \id \lambda \rho_1(\lambda)+{\cal
O}(L^{-1})\ , \label{Eq:particlemom}\\
E^h &= Le - \eps(\lambda^h) +o(1), \qquad \abs{\lambda^h}<B\ , 
\label{Eq:holeen}\\
P^h &= \pi - 2\pi \int_0^{\lambda^h} \id \lambda \rho_1(\lambda)+{\cal
O}(L^{-1})\ , \label{Eq:holemom}
\end{align}
where the \emph{dressed energy} $\epsilon_1(\lambda)$ is a solution to
the linear integral equation
\be
\eps_1(\lambda) = \eps^{(0)}_1(\lambda) - \frac{1}{2\pi} \int_{-B}^B \id \mu\
a_2(\mu-\lambda)\ \eps_1(\mu).
\ee 
The rapidities $\lambda^p$ and $\lambda^h$ are continuous parameters
above and below the ``Fermi-edge'' respectively. They are related to
the parameters $I^p$ and $I^h$ through Takahashi's equations \fr{LBAE}.

\begin{figure}
  \begin{center}
\includegraphics[width=0.8\textwidth]{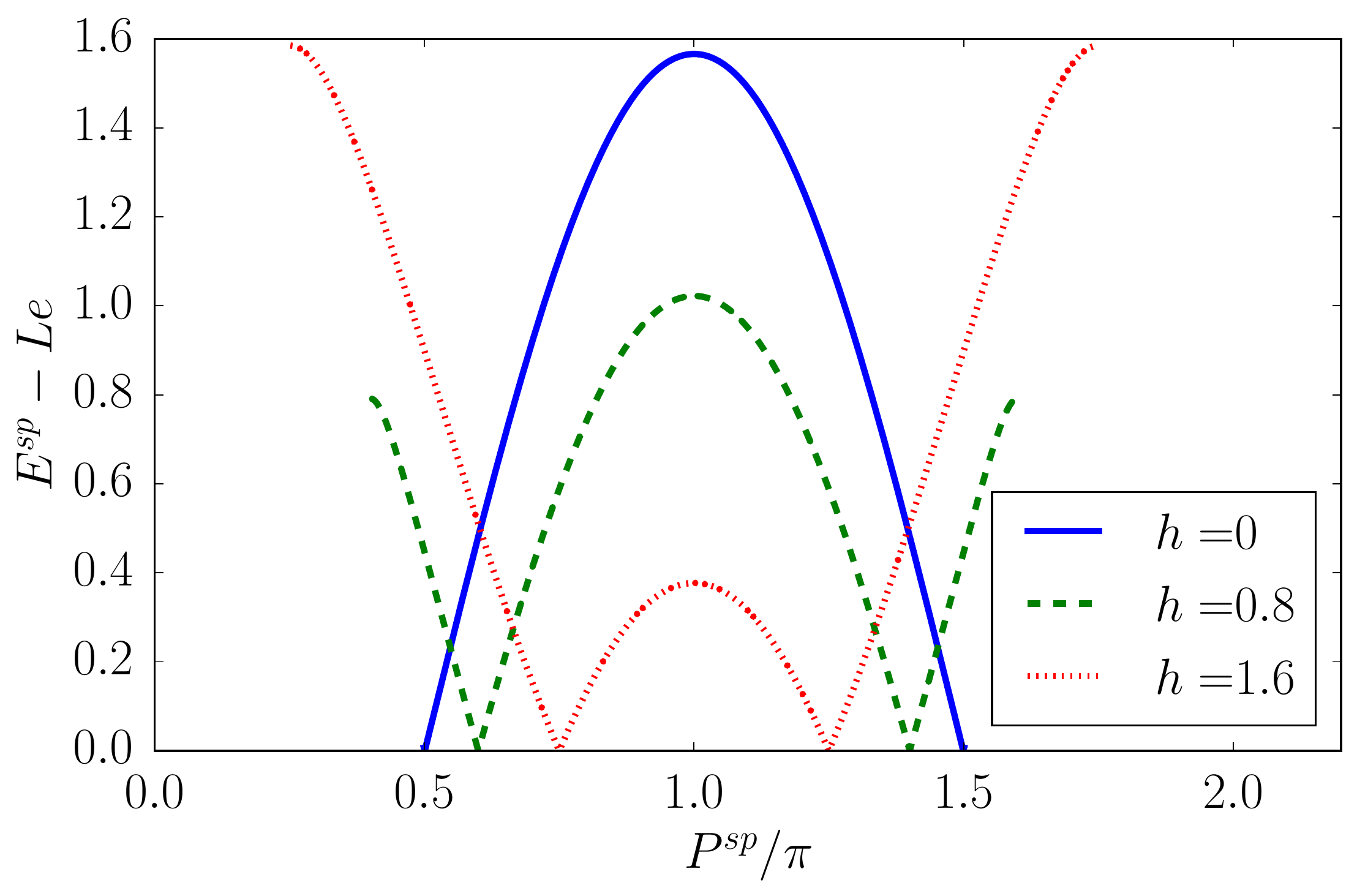}
\end{center}
\caption{Energy-momentum dispersion of the one-parameter excitations for
different values of the magnetic field h.}
\label{Fig:dispdifferenth}
\end{figure}

The excitation energy for a one-spinon excitation can now be
extracted by simply subtracting the extensive part of the energy
(which equals the ground state energy per site of the Heisenberg
chain), which allows us to extract the spinon energy and momentum
\be
\epsilon_s(\lambda)=|\eps_1(\lambda)|\ ,\qquad
p_s(\lambda)=\pi+2\pi\ {\rm sgn}(|\lambda|-B)
\int_0^\lambda d\mu\ \rho_1(\mu)\ .
\label{epss}
\ee
The corresponding dispersion relation is plotted for several values of
magnetic field $h$ in Fig.~\ref{Fig:dispdifferenth} where the value for the 
magnetic field is fixed by imposing $\eps_1(B)=0$.
We note that by construction the spinon dispersion is identical to the
one extracted from the two-spinon excitation of the Heisenberg model
with even chain lengths $L$, apart from a shift in momentum by $\pi$.

\subsection{Excitations involving several particles and/or holes}
As $\delta H$ commutes with $S^z$, the decay of the single-particle (hole)
excitation described above can only involve excited states with the
same $S^z$ quantum number. These are obtained in the following ways:
\begin{enumerate}
\item{} One can consider solutions of Takahashi's equations only
involving 1-strings. These will involve additional particle-hole
excitations on top of the 1-spinon excitation constructed above.
\item{} One can consider solutions of Takahashi's equations involving 
$n$-strings with $n\geq 2$. As a result of the magnetic field these 
excitations have a gap.
\item{} One can consider excitations of the form \fr{descendant} that
are not SU(2) highest-weight states. These again 
have a gap for $h>0$ because
\be
[S^-,H(J,\Delta,h)]=-hS^-
\ee
\end{enumerate}
As we are dealing with an interacting theory, this leaves us with an
infinite number of possible decay channels, i.e. even to first order in
perturbation theory in $\kappa$, a single spinon can decay into
excitations involving $3,5,7,\dots$ particles. As in one dimension 
the accessible phase space shrinks with the number of particles
involved \cite{mussardo}, it is reasonable to assume that the dominant
decay channels will involve excitations with low numbers of particles.
In the following we will focus on excitations involving $3$ particles.
We have considered a class of five-particle excitations where we
excite two particle- and hole-type excitations in addition to the
one-spinon excitation, and found the corresponding decay rate to be 
smaller (see section \ref{sec:decay}).

\subsubsection{``pph-excitation''}

This excitation involves only 1-strings and corresponds to
configurations of the (half-odd) integers $I^1_\alpha$ looking as follows

\begin{figure}[ht]
  \begin{subfigure}{0.49\linewidth}
    \centering
    \includegraphics[width=1.0\linewidth]{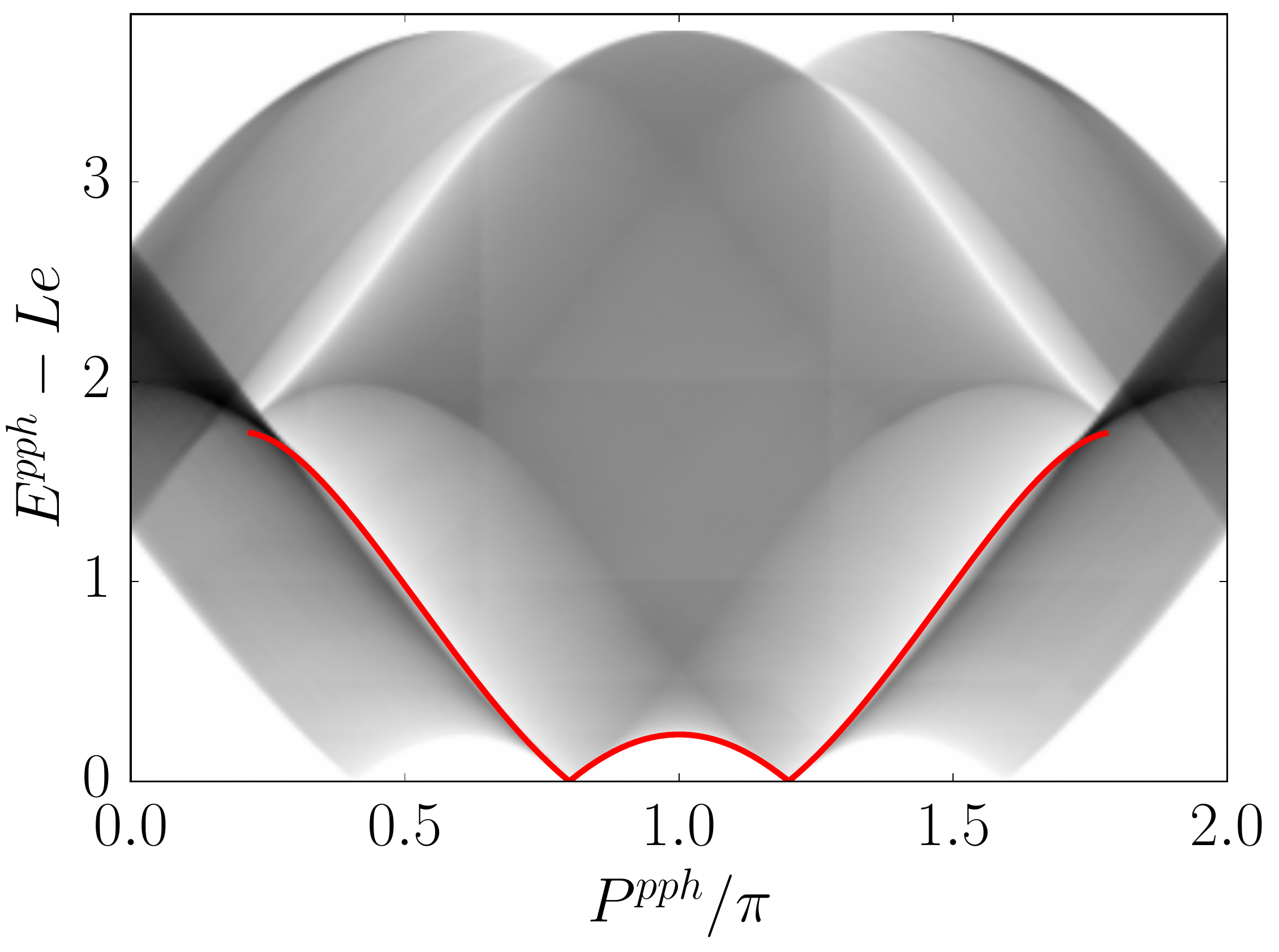} 
  \caption{}
    \label{Fig:pph_enmom}
  \end{subfigure}
  \begin{subfigure}{0.49\linewidth}
    \centering
    \includegraphics[width=1.0\linewidth]{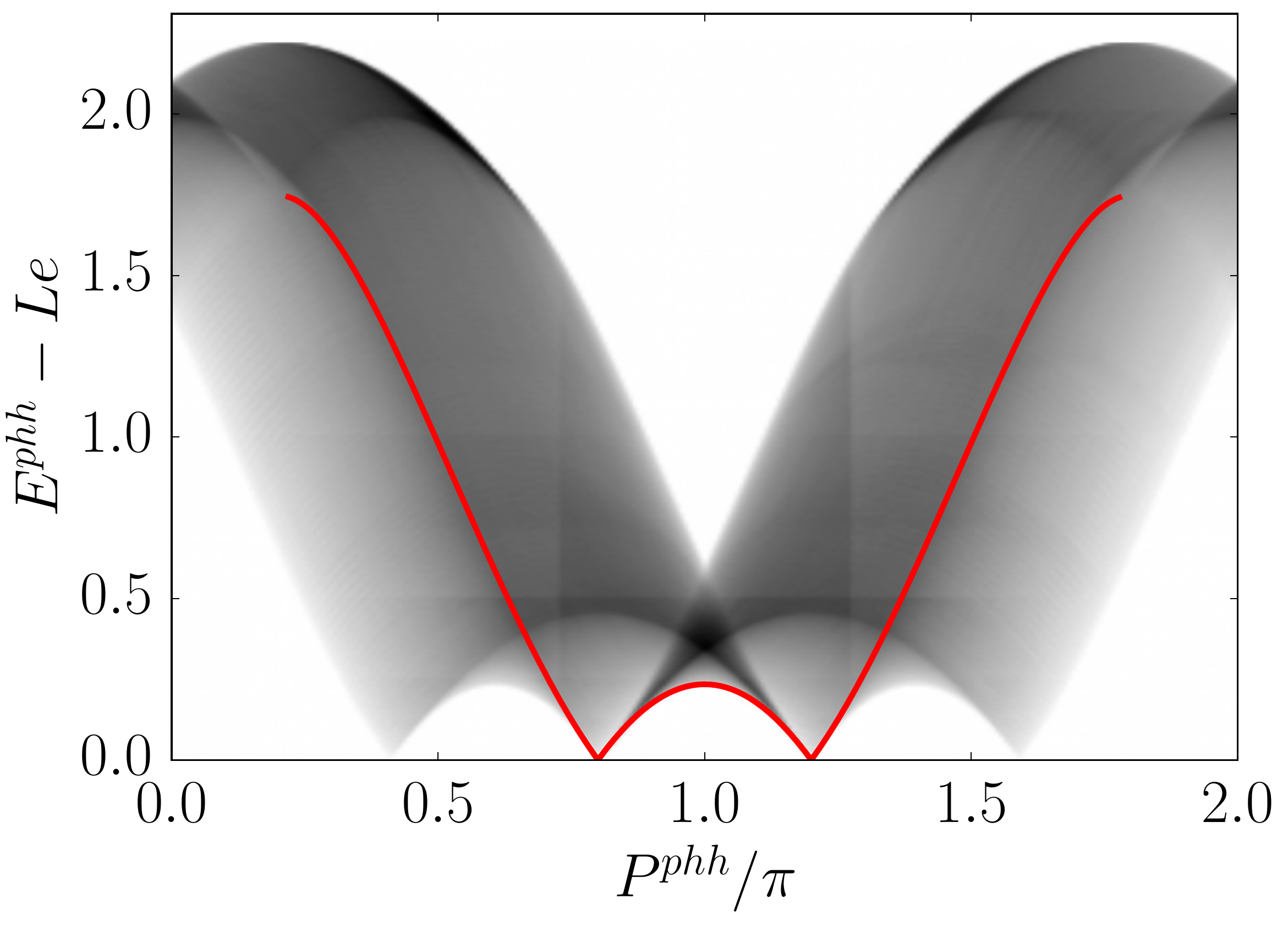} 
    \caption{}
    \label{Fig:phh_enmom}
  \end{subfigure} 
  \begin{subfigure}{0.49\linewidth}
    \centering
    \includegraphics[width=0.99\linewidth]{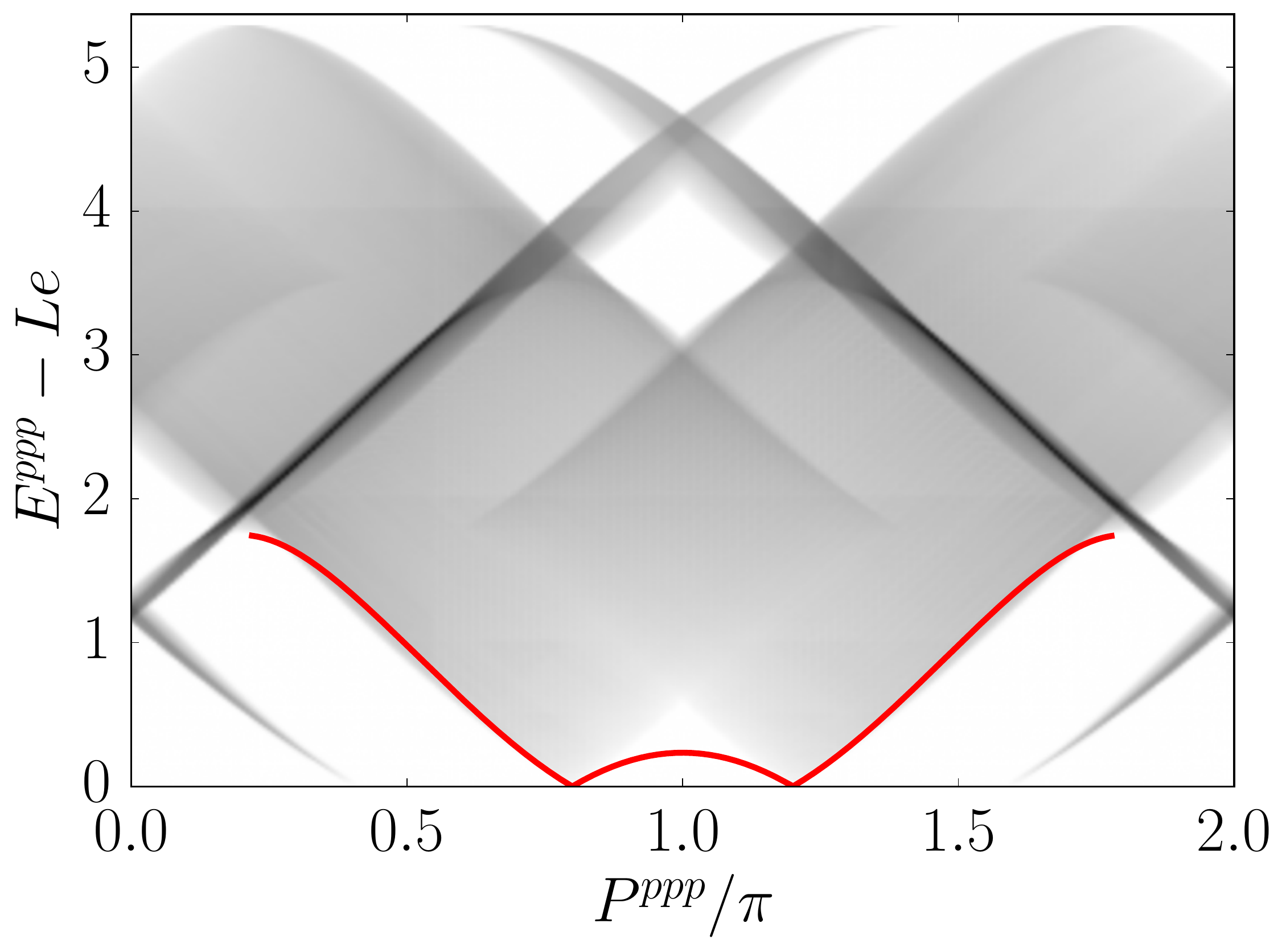} 
    \caption{}
    \label{Fig:ppp_enmom}
  \end{subfigure} 
  \begin{subfigure}{0.49\linewidth}
    \centering
    \includegraphics[width=0.99\linewidth]{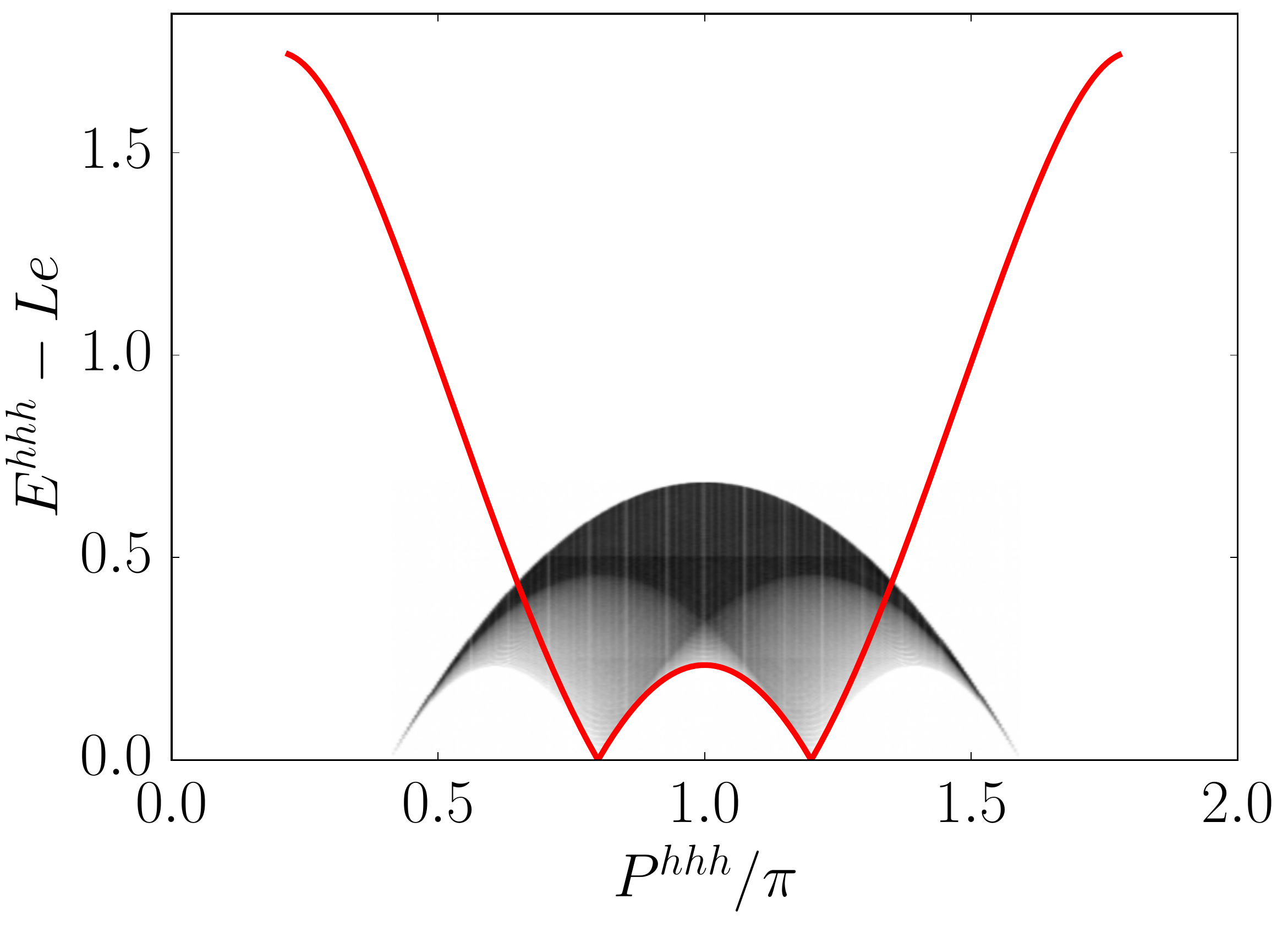} 
    \caption{}
    \label{Fig:hhh_enmom}
  \end{subfigure} 
  \caption{Excitation continua (grey) for (a) two particle one hole (b) one particle two holes (c) three particle and (d) three holes at
magnetization $m=3/10$. The 1-spinon dispersion is shown in red. The
shading of the continuum reflects the density of states (see main
text). Decay of the single spinon is kinematically allowed in part of
the Brillouin zone.
}
\end{figure}

\begin{center}
        \noint
        \border
        \intrange{-9}{10}
        \fillI{-4}{5}
        \holeadd{1}{I^{h}}
        \particleadd{8}{I^{p}_2}
        \particleadd{-6}{I^{p}_1}
        \intLabel{I_\alpha^1}
        \render
\end{center}
States of this kind can be thought of as a sub-class of 3-spinon
excitations that involves two particles and one hole, which are
parametrized by $I^h$ and $I^p_{1,2}$ respectively (or equivalently by
the corresponding rapidities $\lambda^h$, $\lambda^p_{1,2}$).
Energy and momentum of this excitation are given by
\begin{align}
E^{pph} &= Le + \epsilon_s(\lambda_1^{p}) + \epsilon_s(\lambda_2^{p}) 
+ \epsilon_s(\lambda^{h})+o(1)\ ,\\
P^{pph} &= p_s(\lambda^p_1) +p_s(\lambda^p_2) +p_s(\lambda^h)+{\cal
  O}(L^{-1})\ ,
\label{Epph}
\end{align}
where $\epsilon_s(\lambda)$ and $p_s(\lambda)$ are defined in
\fr{epss}. The excitation energy is obtained subtracting the ground
state energy, and the corresponding continuum of 3-spinon excited
states is shown in Fig.~\ref{Fig:pph_enmom}. 
The grey shading reflects the density of excitations at given values of
energy and momentum. Darker regions correspond to higher densities. 
The intensity of the shading is obtained by considering large but
finite $L$ and varying $I^h$ and $I^p_{1,2}$ over all allowed
values for a given excitation, and calculating approximate values of
$\lambda^h,\lambda^p_{1,2}$ by solving the equation
\be
p_s(\lambda^h)=\frac{2\pi I^h}{L}\ ,\quad
p_s(\lambda^p_j)=\frac{2\pi I^p_j}{L}\ ,\ j=1,2.
\ee
The corresponding approximate excitation energy is then obtained by
substituting these values into \fr{Epph}. Each set $\{I^h,I^p_{1,2}\}$
provides one point in the $P^{pph}$-$E^{pph}$-plane and the collection
of all these points generates a shading that reflects the density of
states.

We see that for momenta $\pi
(\frac{1}{2}+m)<p_s(\lambda)<\pi(\frac{3}{2}-m)$ decay of the 1-spinon
excitation is kinematically forbidden, while it is allowed for some
values in the regions $p_s(\lambda)>\pi(\frac{3}{2}-m)$ and
$p_s(\lambda)<\pi(\frac{1}{2}+m)$. 

\subsubsection{``phh-excitation''}
This excitation involves only 1-strings and corresponds to
configurations of the (half-odd) integers $I^1_\alpha$ looking as follows
\begin{center}
        \noint
        \border
        \intrange{-9}{10}
        \fillI{-4}{5}
        \holeadd{1}{I^{h}_2}
        \holeadd{-3}{I^{h}_1}
        \particleadd{9}{I^{p}}
        \particleadd{-5}{}
        \particleadd{6}{}
        \intLabel{I^1_\alpha}
        \render
\end{center}
States of this kind are a sub-class of 3-spinon
excitations that involves one particle and two holes, which are
parametrized by $I^p$ and $I^h_{1,2}$ or equivalently by
the corresponding rapidities $\lambda^p,\lambda^h_{1,2}$.
Energy and momentum of this excitation are given by
\begin{align}
E^{phh} &= Le + \epsilon_s(\lambda^{p}) + \epsilon_s(\lambda^{h}_1)
+ \epsilon_s(\lambda^{h}_2)  +o(1)\ ,\\
P^{phh} &= p_s(\lambda^p) +p_s(\lambda^h_1)+p_s(\lambda^h_2)
+{\cal   O}(L^{-1})\ ,
\end{align}
where $\epsilon_s(\lambda)$ and $p_s(\lambda)$ are defined in
\fr{epss}. 
The excitation energy is again obtained by subtracting the ground
state energy and is shown as a function of the total momentum in
Fig.~\ref{Fig:phh_enmom}.
\subsubsection{``ppp-excitations''}
This excitation involves only 1-strings and corresponds to
configurations of the (half-odd) integers $I^1_\alpha$ looking as follows
\begin{center}
        \noint
        \border
        \intrange{-9}{10}
        \fillI{-4}{5}
        \particleadd{7}{I^{p}_2}
        \particleadd{-8}{I^{p}_1}
        \particleadd{9}{I^{p}_3}
        \holeadd{-4}{}
        \holeadd{5}{}
        \intLabel{I_\alpha^1}
        \render
\end{center}
States of this kind are a sub-class of 3-spinon
excitations that involves three particles.
Energy and momentum of this excitation are
\begin{align}
E^{ppp} &= Le + \sum_{j=1}^3\epsilon_s(\lambda_j^{p})+o(1)\ ,\\
P^{ppp} &= \sum_{j=1}^3p_s(\lambda^p_j) +{\cal O}(L^{-1}),
\end{align}
where $\epsilon_s(\lambda)$ and $p_s(\lambda)$ are defined in
\fr{epss}. 
The excitation energy is again obtained by subtracting the ground
state energy and is shown as a function of the total momentum in
Fig.~\ref{Fig:ppp_enmom}.

\subsubsection{``hhh-excitations''}
This excitation involves only 1-strings and corresponds to
configurations of the (half-odd) integers $I^1_\alpha$ looking as follows
\begin{center}
        \noint
        \border
        \intrange{-9}{10}
        \fillI{-4}{5}
        \holeadd{1}{I^{h}_3}
        \holeadd{-3}{I^{h}_2}
        \holeadd{-4}{I^{h}_1}
        \particleadd{-5}{}
        \particleadd{7}{}
        \particleadd{-6}{}
        \particleadd{6}{}
        \intLabel{I_\alpha^1}
        \render
\end{center}
States of this kind are a sub-class of 3-spinon
excitations that involves three holes. Energy and momentum of this
excitation are 
\begin{align}
E^{hhh} &= Le + \sum_{j=1}^3\epsilon_s(\lambda_j^{h})+o(1)\ ,\\
P^{hhh} &= \sum_{j=1}^3p_s(\lambda^h_j) +{\cal O}(L^{-1}),
\end{align}
where $\epsilon_s(\lambda)$ and $p_s(\lambda)$ are defined in
\fr{epss}. 
The excitation energy is again obtained by subtracting the ground
state energy and is shown as a function of the total momentum in
Fig.~\ref{Fig:hhh_enmom}.

\subsubsection{Excitations involving a single 2-string}
We now turn to the simplest excitation involving a single
2-string. This corresponds to solutions of \fr{LBAE} with $M_1=N-2$,
$M_2=1$ and configurations of the half-odd integers $I^1_\alpha$,
$I^2_1$ of the kind
\begin{center}
        \noint
        \border
        \intrange{-9}{10}
        \fillI{-4}{5}
        \particleadd{8}{I^p}
        \intLabel{I_\alpha^1}
        \render
\end{center}

\begin{figure}[ht] 
  \begin{subfigure}{0.49\linewidth}
    \centering
    \includegraphics[width=1.0\linewidth]{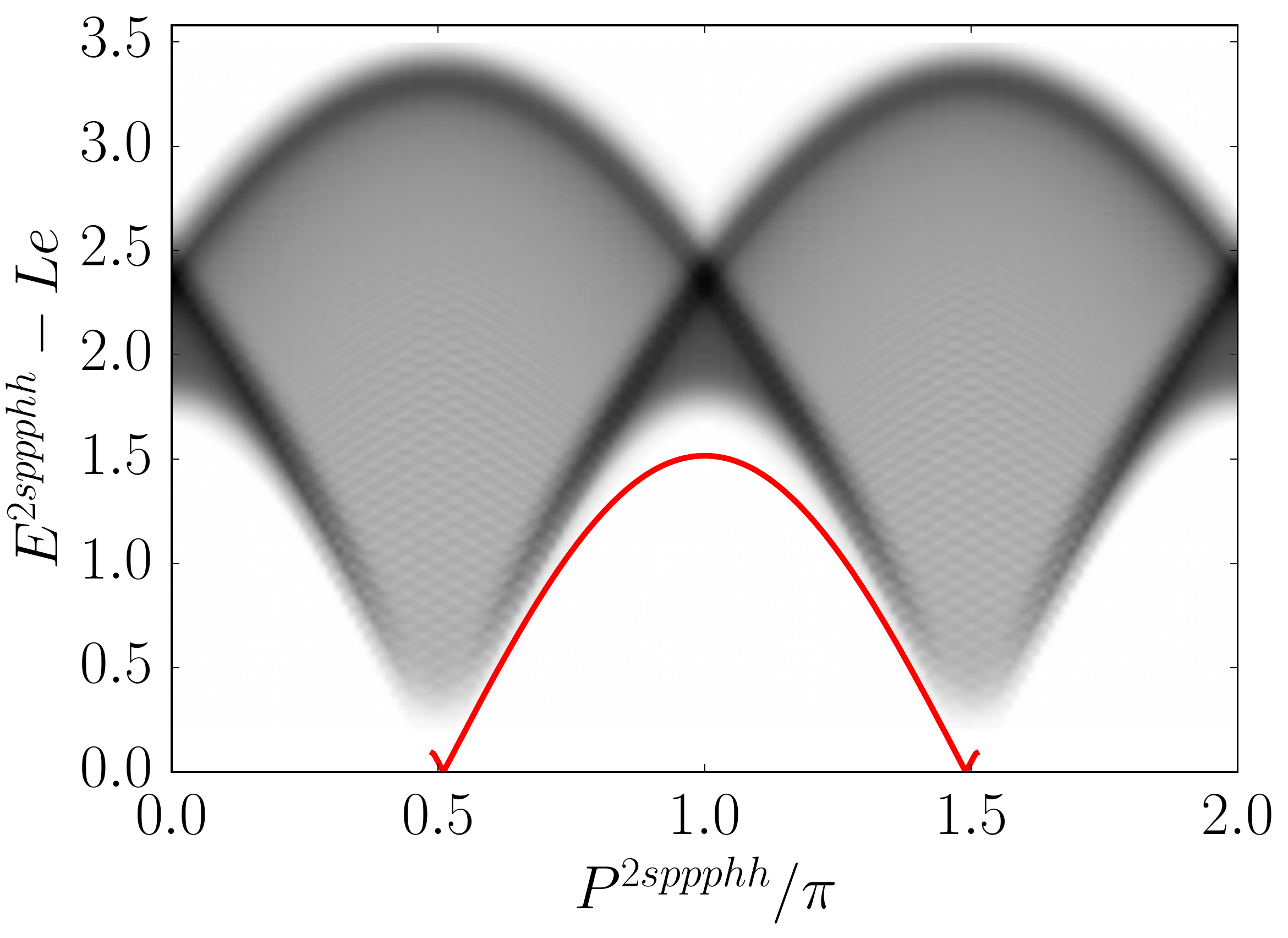} 
  \caption{}
  \end{subfigure}
  \begin{subfigure}{0.49\linewidth}
    \centering
    \includegraphics[width=1.0\linewidth]{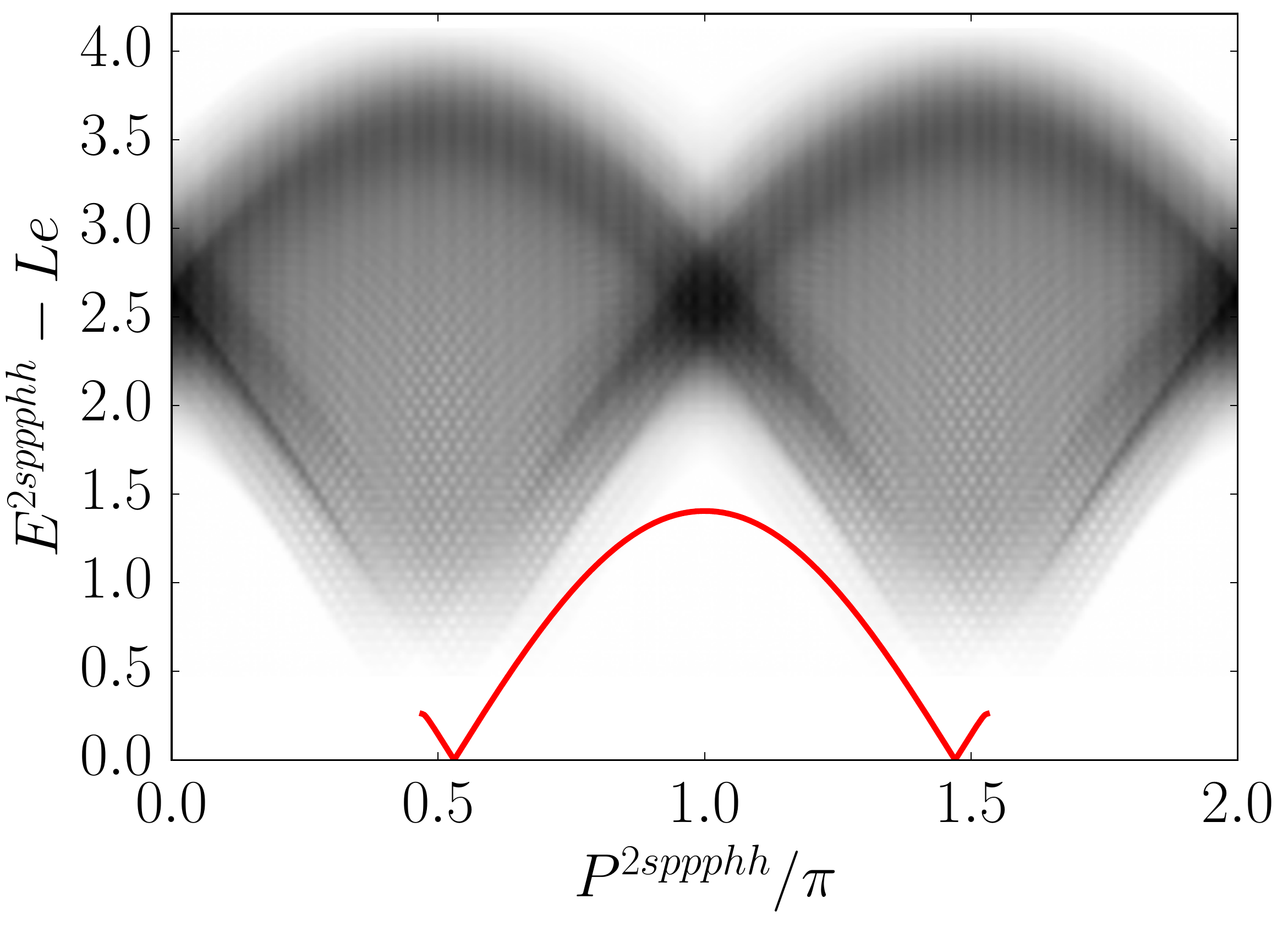} 
    \caption{}
  \end{subfigure} 
  \begin{subfigure}{0.49\linewidth}
    \centering
    \includegraphics[width=0.99\linewidth]{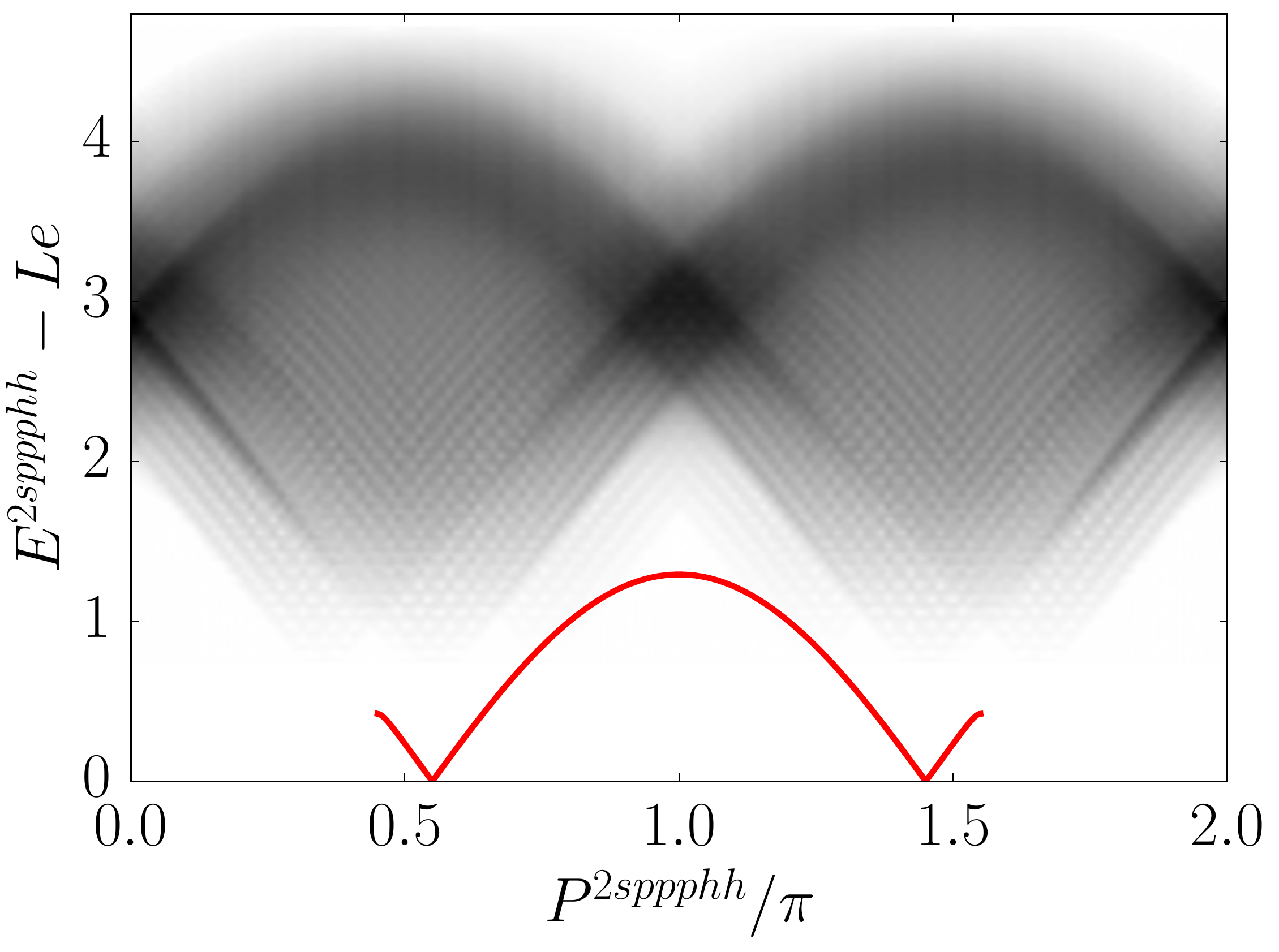} 
    \caption{}
  \end{subfigure} 
  \begin{subfigure}{0.49\linewidth}
    \centering
    \includegraphics[width=0.99\linewidth]{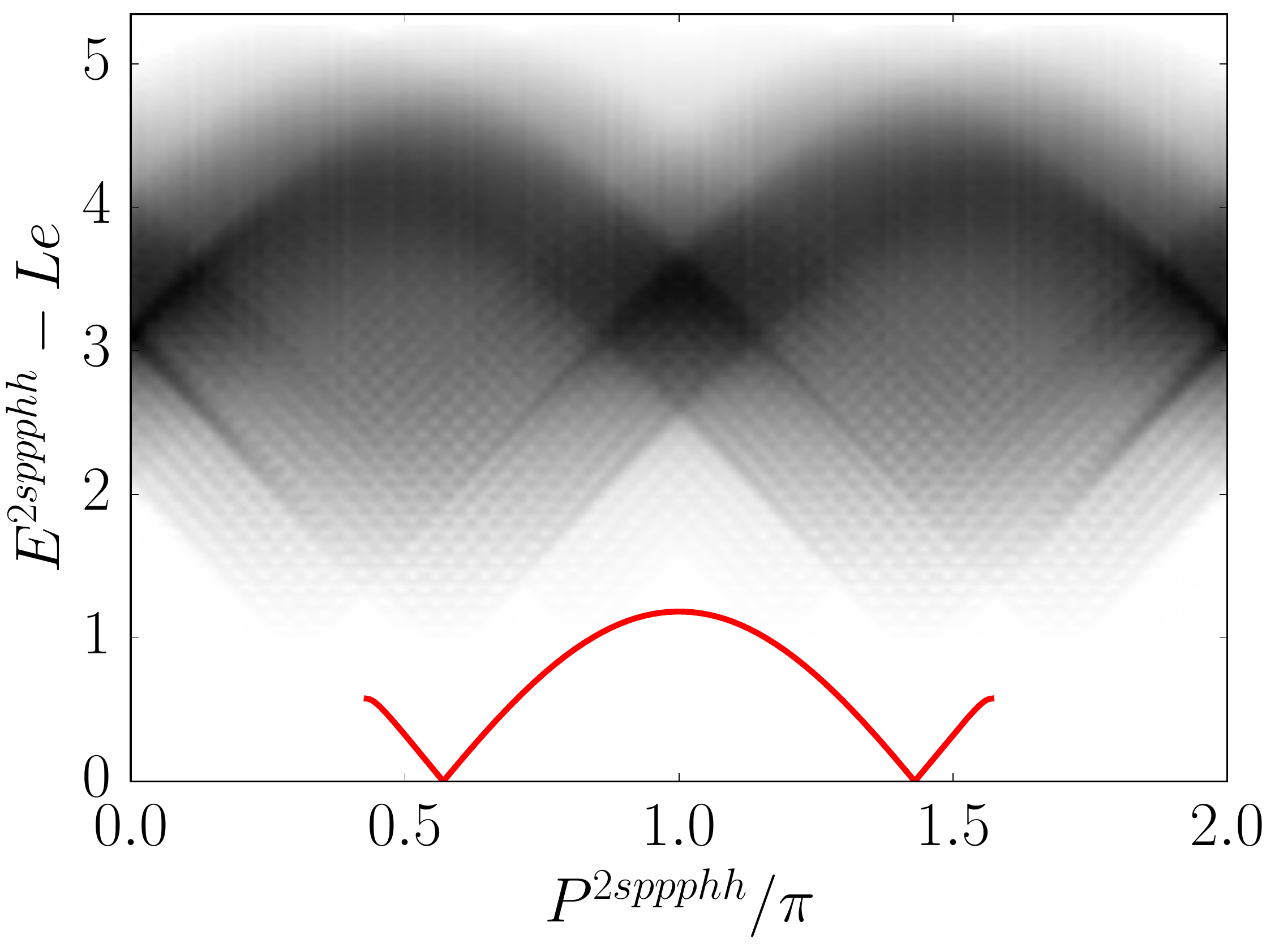} 
    \caption{}
  \end{subfigure} 
\caption{One two-string and $3p2h$ excitation continuum (grey) at
magnetization (a) $m=0.01$, (b) $m=0.03$, (c) $m=0.05$ and (d) $m=0.07$, 
(the corresponding magnetic fields are $h=0.09$, $h=0.26$, $h=0.42$
and $h=0.58$ respectively). The 1-spinon dispersion is shown in
red. The shading of the  
continuum reflects the density of states (see main text). Decay of the 
single spinon is kinematically not allowed.}
\label{Fig:stringsolutionsenergyth}
\end{figure}

\begin{center}
        \noint
        \intrange{-4}{5}
        \fillI{2}{2}
        \intLabel{I^2_1}
        \render
\end{center}
We note that the permitted values for $I^2_1$ have range
\be
|I^2_1|\leq \frac{1}{2}\left[L-2N\right].
\ee
The excitation is parametrized by the two half-odd integers $I^p,
I^2_1$ or equivalently the corresponding rapidities $\lambda^p,\lambda^s$.
Energy and momentum of this excitation are given by
\begin{align}
E^{2sp} &= Le+ \epsilon_s(\lambda^p)+\eps_2(\lambda^s) \, , \qquad
\abs{\lambda^p}>B\ , \\ 
P^{2sp} &= p_s(\lambda^p) +  p_2(\lambda^s)\ ,
\label{2sp}
\end{align}
where $\eps_2$ and $p_2(\lambda)$ are given by \cite{takahashi}
\bea
\eps_2(\lambda) &=& h+\int_{|\mu|>B} \id \mu
a_1(\lambda-\mu)\ \eps_1(\mu) \ ,\nn
p_s(\lambda)&=&\theta\Big(\frac{\lambda}{2}\Big)-\int_{-B}^Bd\mu\
\theta_{21}(\lambda-\mu)\ \rho_1(\mu).
\eea
Excitation continua that encompass the two-particle continuum \fr{2sp}
are obtained by adding particle-hole excitations, e.g.
\begin{align}
E^{2s3p2h} &= Le+ \sum_{j=1}^3\epsilon_s(\lambda^p_j)+
\sum_{k=1}^2\epsilon_s(\lambda^h_k)+\eps_2(\lambda^s) \, , \qquad
\abs{\lambda^h_j}<B<\abs{\lambda^p_k}\ , \\ 
P^{2s3p2h} &= \sum_{j=1}^3p_s(\lambda^p_j)+
\sum_{k=1}^2p_s(\lambda^h_k)+p_2(\lambda^s) \ .
\label{2s3p2h}
\end{align}
The continuum \fr{2s3p2h} is shown in
Fig.~\ref{Fig:stringsolutionsenergyth} for several magnetizations. We
see that the single spinon excitation cannot decay into the 2-string
excitation for kinematic reasons.  
If we keep on adding particle-hole excitation at small magnetic fields
decay of the 1-spinon excitation will eventually become kinematically
allowed. However, the decay rate is then expected to be negligible on
the basis of aforementioned phase-space arguments, \emph{cf.}
Ref.~\cite{mussardo}. 
\subsubsection{Excitations involving longer strings}
Excitations involving longer strings have larger gaps at finite
magnetic fields \cite{takahashi}. We expect contributions from decay
channels involving such excitations to be small for the same reasons
we put forward in the 2-string case above.

\subsubsection{Excitations that are not highest weight states}
As mentioned above, excitations which are not highest weight states have
gaps that are proportional to the magnetic field $h$. Nevertheless, decay of
a single spinon into excitations that are not highest weight states
will generally be allowed at sufficiently small $h$. As an example let
us consider highest-weight states with $M_1=N-1$, $M_{n\geq 2}=0$. The
lowest energy states in this sector correspond to 
integers 
\be
I^1_\alpha=-\frac{N}{2}+\alpha\ ,\quad\alpha=1,\dots,N,
\ee
or
\be
I^1_\alpha=-\frac{N}{2}-1+\alpha\ ,\quad\alpha=1,\dots,N,
\ee
\begin{center}
        \integers
        \border
        \intrange{-9}{9}
        \fillI{-4}{4}
        \particleadd{5}{}
        \intLabel{I_\alpha^1}
        \render
\end{center}
\begin{center}
        \integers
        \border
        \intrange{-9}{9}
        \fillI{-4}{4}
        \particleadd{-5}{}
        \intLabel{I_\alpha^1}
        \render
\end{center}
In complete analogy to our discussion above, these can be viewed as
particular limits of a 1-spinon excitation. Acting with the spin
lowering operator gives a 1-parameter excited state with a dispersion
that equals the 1-spinon dispersion shifted upwards in energy by
$h$. Hence decay of our 1-spinon excitation into this particular
descendant state is kinematically not allowed. However, if we add an
additional particle-hole pair the situation changes. Let us consider
configurations of integers such as
\begin{center}
        \noint
        \border
        \intrange{-9}{9}
        \fillI{-4}{4}
        \holeadd{1}{I^{h}}
        \particleadd{8}{I^{p}_2}
        \particleadd{-6}{I^{p}_1}
        \intLabel{I_\alpha^1}
        \render
\end{center}
States of this kind can be thought of as a sub-class of 3-spinons.
Energy and momentum of the excitation obtained by acting with the
spin-lowering operator on this state are given by
\begin{align}
E^{pph}_{\rm desc} &= Le + \epsilon_s(\lambda_1^{p}) + \epsilon_s(\lambda_2^{p}) 
+ \epsilon_s(\lambda^{h})+h+o(1)\ ,\\
P^{pph}_{\rm desc} &= \pi + p_s(\lambda^p_1) +p_s(\lambda^p_2) +p_s(\lambda^h)+{\cal
  O}(L^{-1})\ .
\label{Epphdesc}
\end{align}
Inspection of Fig.~\ref{Fig:pph_enmom} shows that decay of the
1-spinon excitation into this continuum is kinematically allowed at
sufficiently weak fields.  However, as shown in \ref{Appendix:Descendants}
this decay is strongly suppressed for large system sizes $L$.

\section{Algebraic Bethe Ansatz and Derivation of the Matrix Element}
\label{sec:matrixel}

\subsection{Algebraic Bethe Ansatz}
\label{ssec:ABA}
In order to determine decay rates we require matrix elements of the
perturbing operator between energy eigenstates. These can be obtained
using the Algebraic Bethe Ansatz \cite{korepin}. In the following we
will first consider the XXZ case with anisotropy parameter
$\Delta=\cos\gamma$ and only later specialize to the isotropic limit $\Delta=1$.
A key object is the monodromy matrix 
\begin{align}
T(\lambda)=
\begin{pmatrix} A(\lambda) & B(\lambda) \\ C(\lambda) & D(\lambda)
\end{pmatrix}\ ,
\end{align}
where $A,B,C,D$ are operators acting on the Hilbert space of the chain
and $\lambda$ is known as spectral parameter. The mondromy matrix fulfils
the Yang-Baxter equation
\begin{align}
R(\lambda-\mu)(T(\lambda)\otimes 1)&(1\otimes T(\mu))=\notag \\
& (1\otimes T(\mu))(T(\lambda)\otimes 1) R(\lambda-\mu)\ ,
\end{align}
where the $R$-matrix has the form
\begin{align}
R(\lambda,\mu)=\begin{pmatrix}
1 & 0 & 0 & 0 \\
0 & b(\lambda,\mu) & c(\lambda,\mu) & 0 \\
0 & c(\lambda,\mu) & b(\lambda,\mu) & 0 \\
0 & 0 & 0 & 1
\end{pmatrix}.
\end{align}
Here we have defined
\begin{align}
b(\lambda)&=\frac{\sinh(\lambda)}{\sinh(\lambda+i\gamma)} \\
c(\lambda)&=\frac{\sinh(i\gamma)}{\sinh(\lambda+i\gamma)}
\end{align}
The Yang-Baxter algebra determines intertwining relations for the
operators $A,B,C,D$. Eigenstates of \eqref{XXZhamiltonian} can be constructed
as
\begin{align}
\ket{\lambda_1,\dots,\lambda_N}=\prod_{i=1}^N B(\lambda_i)
\ket{0}\ ,
\label{eigenstates}
\end{align}
where the set of rapidities $\{\lambda_i\}_{i\in \{1,\dots,N\}}$ are
solutions to the Bethe equations
\be
\frac{a(\mu_j)}{d(\mu_j)}=\prod_{k\neq
  j}\frac{b(\mu_k-\mu_j)}{b(\mu_j-\mu_k)}\ ,\quad
j=1,\dots,N.
\label{BAE}
\ee
The reference state $\ket{0}$ satisfies
\begin{align}
A(\lambda)\ket{0}=\ket{0} & &D(\lambda)\ket{0}=d(\lambda)\ket{0} & &
C(\lambda)\ket{0}=0\ . 
\end{align}
The functions $a(\mu)$ and $d(\mu)$ are given by
\begin{align}
&a(\mu)=1 &d(\mu)=\Big( b(\mu-i\frac{\gamma}{2})\Big)^L .
\end{align}
The isotropic limit $\Delta=1$ corresponds to taking $\gamma\to 0$,
while rescaling the spectral parameters 
\be
\mu_j=\gamma\lambda_j\ ,\quad \lambda_j\ {\rm fixed}.
\label{XXXlimit}
\ee
This recovers the Bethe Ansatz equations
\fr{Bethesinh} from \fr{BAE}. The global spin lowering operator is
obtained as \cite{FT}
\be
-i\lim_{\lambda\to\infty}\lambda \, B(\lambda)=S^-.
\ee

\subsection{Determinant Formulas for Matrix Elements in the XXZ chain}
The Algebraic Bethe Ansatz provides a convenient setting for calculating
scalar products as well as the norm of Bethe states 
\cite{gaudinnorm,korepinnorm,slavnovscalar}. Matrix elements can be
analyzed by utilizing the expression of local spin operators
$\sigma_j^z$ in terms of the operators $A,B,C,D$, \emph{cf.}
Ref.~\cite{Kitanine}. With the help of these relations matrix elements
of spin operators $S^\alpha_j$ between eigenstates of the XXZ
Hamiltonian were derived in Ref.~\cite{Kitanine}, and general operators 
were considered in Ref.~\cite{Kitaninegeneral}. Explicit expressions
for the operator $S^z_jS^z_{j+1}$ were obtained in Ref.~\cite{Klauser}.  
Following the derivation of Ref.~\cite{Klauser} we obtain (see
\ref{Appendix} for details) 
\begin{align}
  &\sum_j\bra{\lambda_1,\dots,\lambda_N} S_j^z S_{j+2}^z 
  \ket{\mu_1,\dots,\mu_M} = \notag
\\ &Le^{iP_{\{\lambda\}}+2iP_{\{\mu\}}}
\delta_{P_{\{\lambda\}},P_{\{\mu\}}} \delta_{M,N} \left\{-\phi(i\gamma)
\frac{\prod_k\phi(\lambda_k+i\gamma/2)^2\phi(\mu_k+i\gamma/2)\phi(\mu_k-i\gamma/2)^{-3}}{\prod_{a>b}\phi(\lambda_b-\lambda_a)\prod_{b>a}\phi(\mu_b-\mu_a)}\right. \notag
\\ &\times\sum_n A_n
\left(\det{G_n^{(1)}+B_n^{(1)}}+\det{G_n^{(2)}+B_n^{(2)}}-\det{G^{(1)}}-\det{G^{(2)}}\right)\notag
\\ & \hspace{1cm}-\frac{1}{2}
\frac{\prod_k\phi(\lambda_k+i\gamma/2)^3\phi(\mu_k-i\gamma/2)^{-3}}{\prod_{a>b}\phi(\lambda_b-\lambda_a)\prod_{a<b}\phi(\mu_b-\mu_a)}
\notag \\ & \left.\hspace{2cm} \times\sum_{n=1} \sum_{m\neq n} A_{nm}
\left[\det{G_{nm}+B_{nm}}-\det{G_{nm}}\right]\right\}\ .
\end{align}
Here $P_{\{\lambda\}}$ is the total momentum of the state
parametrized by the rapidities $\{\lambda_j\}$
\be
P_{\{\lambda\}}=\sum_{j=1}^N\left[\pi+\ln\left(\frac{\lambda_j+\frac{i\gamma}{2}}
{\lambda_j-\frac{i\gamma}{2}}\right)\right],
\ee
and
\begin{align}
  H_{ab}&=\frac{\phi(i\gamma)}{\phi(\lambda_a-\mu_b)}\Big(
a(\mu_b)\prod_{k\neq a} \phi(\lambda_k-\mu_b+i\gamma)-d(\mu_b)\prod_{k\neq a} \phi(\lambda_k-\mu_b-i\gamma)\Big), \\
  (G^{(1)}_n)_{ab} &= \begin{cases} H_{ab} & b\neq n
    \\ \frac{\phi(i\gamma)\phi(2\lambda_a)}{\phi(\lambda_a-\frac{i\gamma}{2})^2\phi(\lambda_a+\frac{i\gamma}{2})^2}
    & b=n \end{cases} \ ,\\
  (G^{(2)}_n)_{ab} &= \begin{cases} H_{ab} & b\neq n \\ \pd{2}{}{x}
    \left(
    \frac{\phi(i\gamma)}{\phi(\lambda_a-x)\phi(\lambda_a-x+i\gamma)}\right)\at{x=\frac{i\gamma}{2}}
    & b=n \end{cases} \ ,\\
  (B^{(1)}_n)_{ab} &= (1-\delta_{bn}) d(\mu_b) \prod_{i\neq n} \phi(\mu_i-\mu_b+i\gamma)\phi(\mu_b+i\gamma/2) \notag \\
        &\times\left( \frac{\phi^\prime(i\gamma)}{\phi(i\gamma)}-\sum_{d\neq n} \frac{\phi(i\gamma)}{\phi(\mu_d-i\gamma/2)\phi(\mu_d+i\gamma/2)}+\sum_{d\neq m,n} \frac{\phi^\prime(\mu_d-i\gamma/2)}{\phi(\mu_d-i\gamma/2)} \right. \notag \\ 
        &\left.-\sum_b
  \frac{\phi^\prime(\lambda_b+i\gamma/2)}{\phi(\lambda_b+i\gamma/2)}+\frac{\phi(i\gamma)}{\phi(\mu_m+i\gamma/2)\phi(\mu_m-i\gamma/2)}\right)
  \ ,\\
  (B^{(2)}_n)_{ab} &= (1-\delta_{bn}) \frac{1}{2} d(\mu_b) 
\prod_{i\neq n}
\frac{\phi(\mu_i-\mu_b-i\gamma)\phi(\mu_b+i\gamma)\phi(i\gamma)}{\phi(\lambda_a+i\gamma/2)\phi(\lambda_a-i\gamma/2)}\ .
  \end{align}
\begin{align}
  (G_{nm})_{ab}&=\begin{cases} H_{ab} & b\neq m,n \\ \frac{\phi(i\gamma)\phi(2\lambda_a)}{\phi(\lambda_a-\frac{i\gamma}{2})^2\phi(\lambda_a+\frac{i\gamma}{2})^2} & b=m  \\ \pd{2}{}{x} \left( \frac{\phi(i\gamma)}{\phi(\lambda_a-x)\phi(\lambda_a-x+i\gamma)}\right)\at{x=\frac{i\gamma}{2}} & b=n \end{cases} \\ 
    (B_{nm})_{ab}&=(1-\delta_{bm})(1-\delta_{bn})d(\mu_b) 
\prod_{i\neq m,n}\frac{\phi(\mu_i-\mu_b-i\gamma)
  \phi(\mu_b+i\gamma/2)^2\phi(i\gamma)}{\phi(\lambda_a-i\gamma/2)\phi(\lambda_a+i\gamma/2)}\ .
\end{align}
Finally, the function $\phi(\lambda)$ is given by
\be
\phi(\lambda)= \begin{cases} \lambda & \Delta=1 \\ \sinh(\lambda) & \abs{\Delta}<1 \end{cases} .
\ee
In the isotropic case $\Delta=1$ of interest to us the matrix element
in the rescaled rapidity variables \fr{XXXlimit} is obtained by simply
setting $\gamma=1$ in the above expressions.

\section{Decay rates}
\label{sec:decay}
\begin{figure}[ht]
  \begin{center}
    \includegraphics[width=0.7\textwidth]{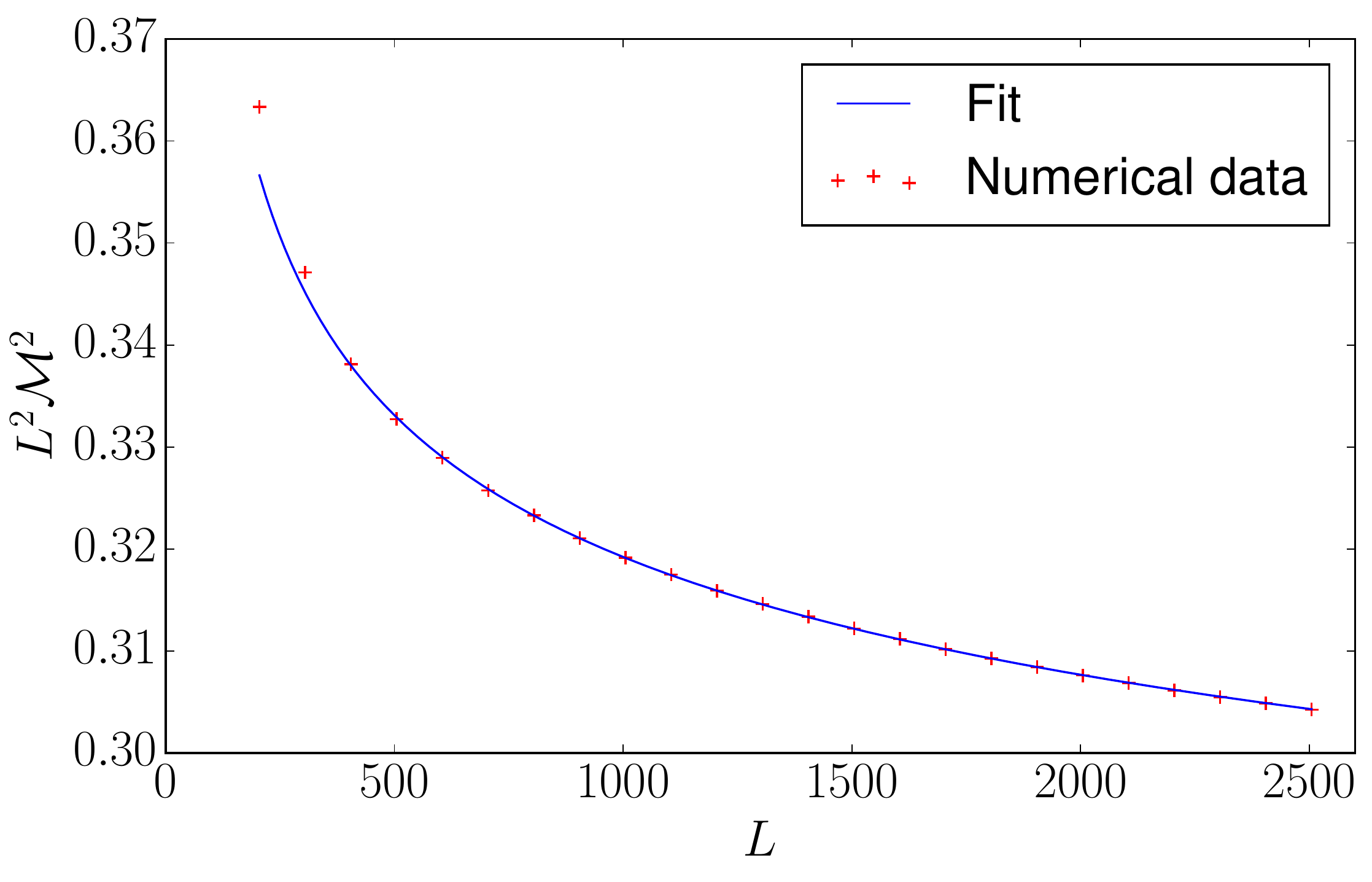}
  \end{center}
\caption{
Scaling of matrix elements with system size $L$. The initial
state contains one spinon with momentum $p=5.5$, while the final state
contains three high-energy excitations with momenta $p_p=4.5, p_p=4.65, p_h=3.65$. 
The fit is to the functional form
(\ref{scalingM}) with $a=0.42$, $b=5.6$, $c=0.044$.
We have also considered additional excitations
around the Fermi points in the final state as well as different momenta and found similar behaviour.
}
\label{Fig:logcorrections}
\end{figure}

We are now in a position to compute the rates of decay of the one-spinon
excitation into the various multi-particle excitations considered
above. In practice the calculation is carried out in a large, finite
volume $L$. The energy eigenstates are of the form \fr{eigenstates} and
involve $N$ rapidity variables, which constitute a solution to the
Bethe Ansatz equations. We will denote the states
corresponding to the one-spinon and the two-particle one-hole continuum by
\be
|N; J\rangle ,\qquad |N;I^p_1,I^p_2,I^h\rangle\ ,
\ee
Our notations for the respective energies \fr{EP} are
\be
E(J)\ ,\quad E^{pph}(I^p_1,I^p_2,I^h)\ .
\ee
Here $J$ and $I^p_{1,2},I^h$ denote the half-odd integers
corresponding to the spinon and the particles/hole respectively. Other
excitations are labelled analogously. The decay rate is then given by
\bea
\fl
\Gamma_{{\rm sp}\rightarrow {\rm
    pph}}(p_J)\equiv\kappa^2 \gamma_{{\rm sp}\rightarrow {\rm  pph}}(p_J)
&=&\pi\kappa^2\sum_{I^p_1,I^p_2,I^h}
|\langle N;I^p_1,I^p_2,I^h|\sum_{j=1}^LS^z_jS^z_{j+2}|N;J\rangle|^2\nn
\fl
&&\quad\times\ \delta_{J,I^p_1+I^p_2-I^h}\ \delta(E^{pph}(I^p_1,I^p_2,I^h)-E(J))\ ,
\label{decayrate}
\eea
where we have used \fr{EP} to simplify the momentum conservation
delta function. The momentum $p_J$ of the initial spinon excitation is
given by 
\be
p_J=
\begin{cases}
\frac{2\pi J}{L} & \text{if } |J|>\frac{N}{2}-1\ ,\\
-\frac{2\pi J}{L} & \text{if } |J|\leq\frac{N}{2}.
\end{cases}
\ee
We regularize the delta function expressing energy conservation by
\be
\delta_\eta(x)=\frac{1}{\sqrt{\pi}\eta}e^{-\left(\frac{x}{\eta}\right)^2},
\ee
where $\lim_{\eta\to 0} \delta_\eta(x)=\delta(x)$. For very small
$\eta$, but still with a sufficient number of final states in the
regime where $\delta_\eta(x)$ is large, we expect the result to be
close to the answer in the thermodynamic limit. 
For \fr{decayrate} to be finite in the thermodynamic limit, the
matrix elements should scale as $L^{-1}$. As shown in
Fig.~\ref{Fig:logcorrections}, the decay in $L$ is very slightly
faster than $L^{-2}$ and is compatible with the functional form
\begin{align}
  \left(\mathcal{M}_{pph}^2 L^2\right) &= \frac{a+b/L}{L^c}\ ,
\label{scalingM}
\end{align}
where $c$ is a very small exponent. In the range of lattice lengths
accessible to us, equally good fits can be obtained by replacing $L^c$ by
$(\ln L)^c$ in (\ref{scalingM}).
The situation is analogous to that for the dynamical structure
factor \cite{Abacus,Pereira,HagemansPaper,shashi1,shashi2,kitaninethermo1,kitaninethermo2,kitanineformfactor1}. For the latter it was shown that in
order to obtain finite results in the thermodynamic limit, an infinite
summation over states that contain additional particle-hole pairs
located at the ``Fermi points'' $\pm B$ was required. On the other
hand, the result obtained by working at a fixed value of $L\approx
1000$ and not carrying out this summation was found to give an
excellent approximation to the thermodynamic limit. We expect the
decay rate to behave in an analogous way. In the following we
determine the contributions of the 3-particle excitations described in
section \ref{sec:excitations} to the decay rate for finite system
sizes in the range $L\sim 500-1000$. We have verified that taking into
account states with one additional particle-hole excitation gives only
small corrections.

\begin{figure}[ht]
  \begin{subfigure}{0.49\linewidth}
    \centering
    \includegraphics[width=1.0\linewidth]{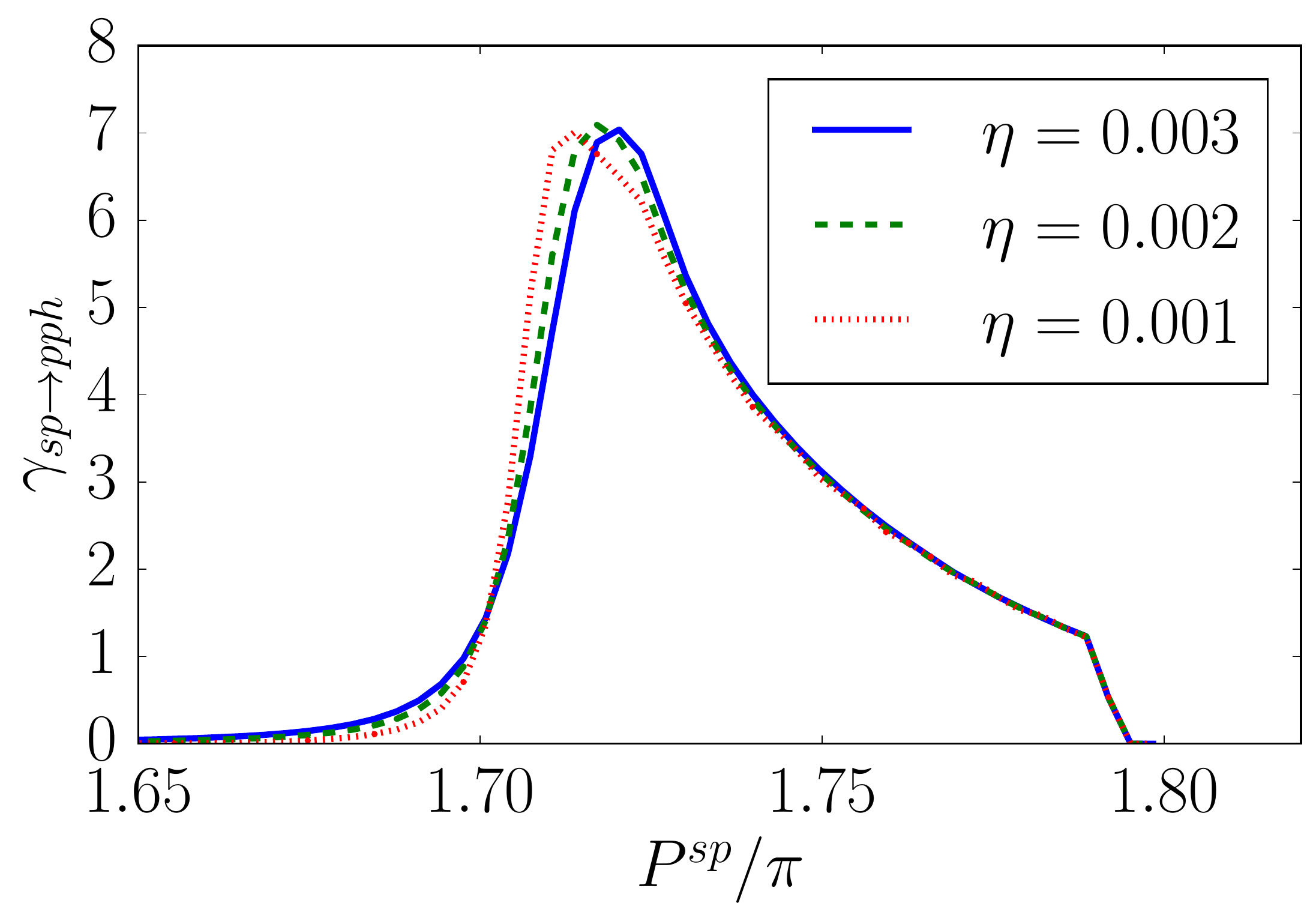} 
  \caption{}
    \label{Fig:deltadecaypph}
  \end{subfigure}
  \begin{subfigure}{0.49\linewidth}
    \centering
    \includegraphics[width=1.0\linewidth]{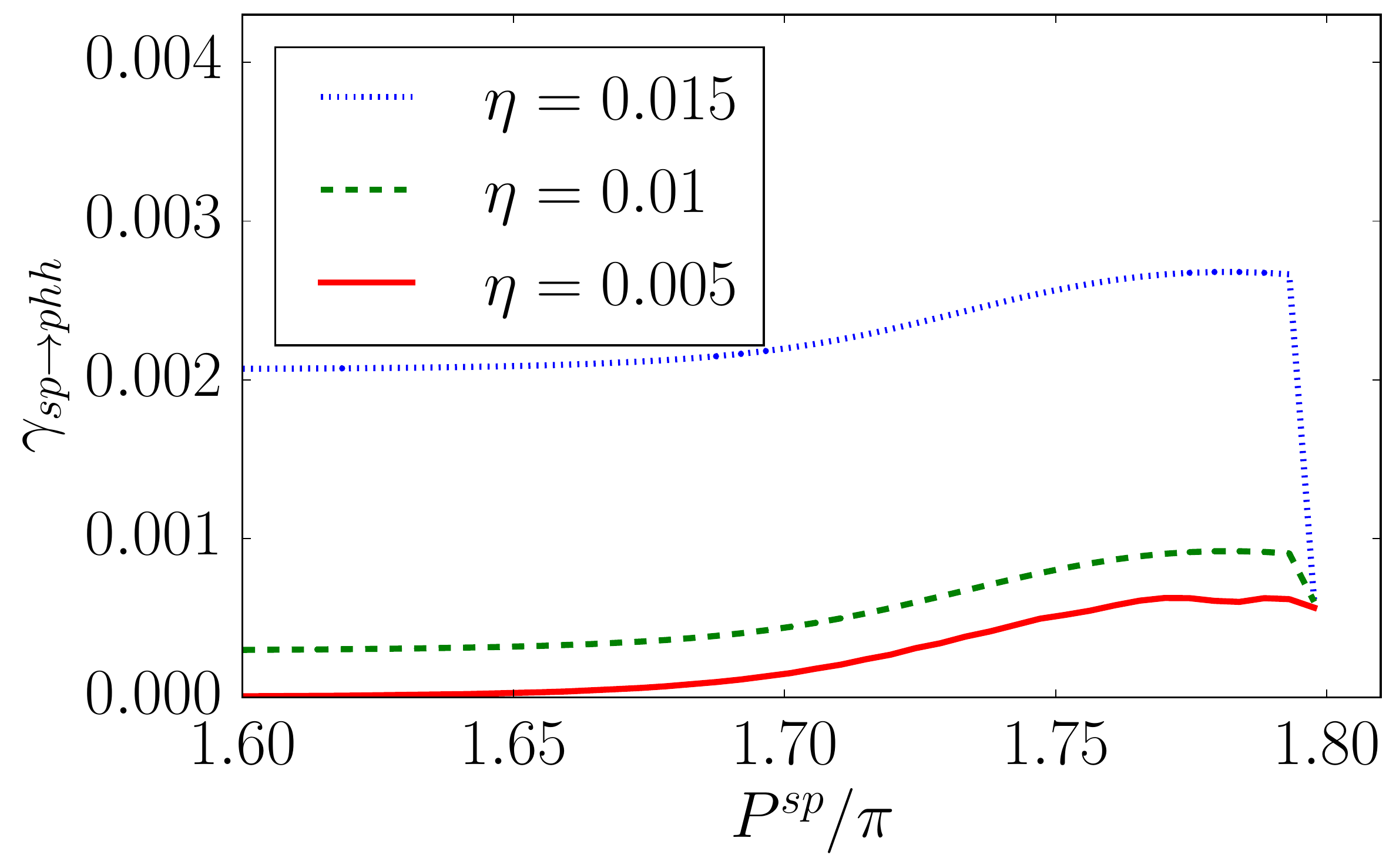} 
    \caption{}
    \label{Fig:deltadecayphh}
  \end{subfigure} 
  \begin{subfigure}{0.49\linewidth}
    \centering
    \includegraphics[width=0.99\linewidth]{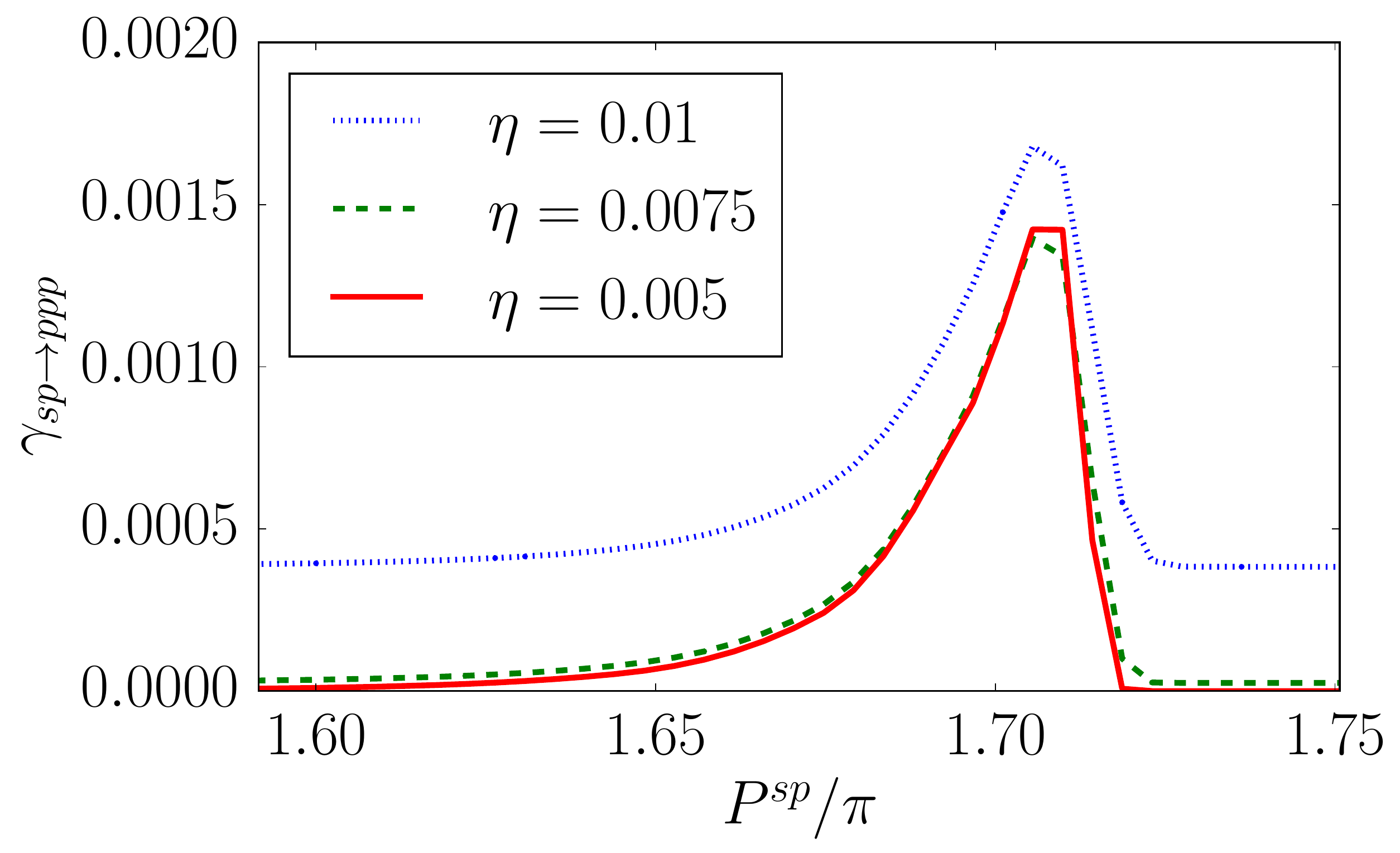} 
    \caption{}
    \label{Fig:deltadecayppp}
  \end{subfigure} 
  \begin{subfigure}{0.49\linewidth}
    \centering
    \includegraphics[width=0.99\linewidth]{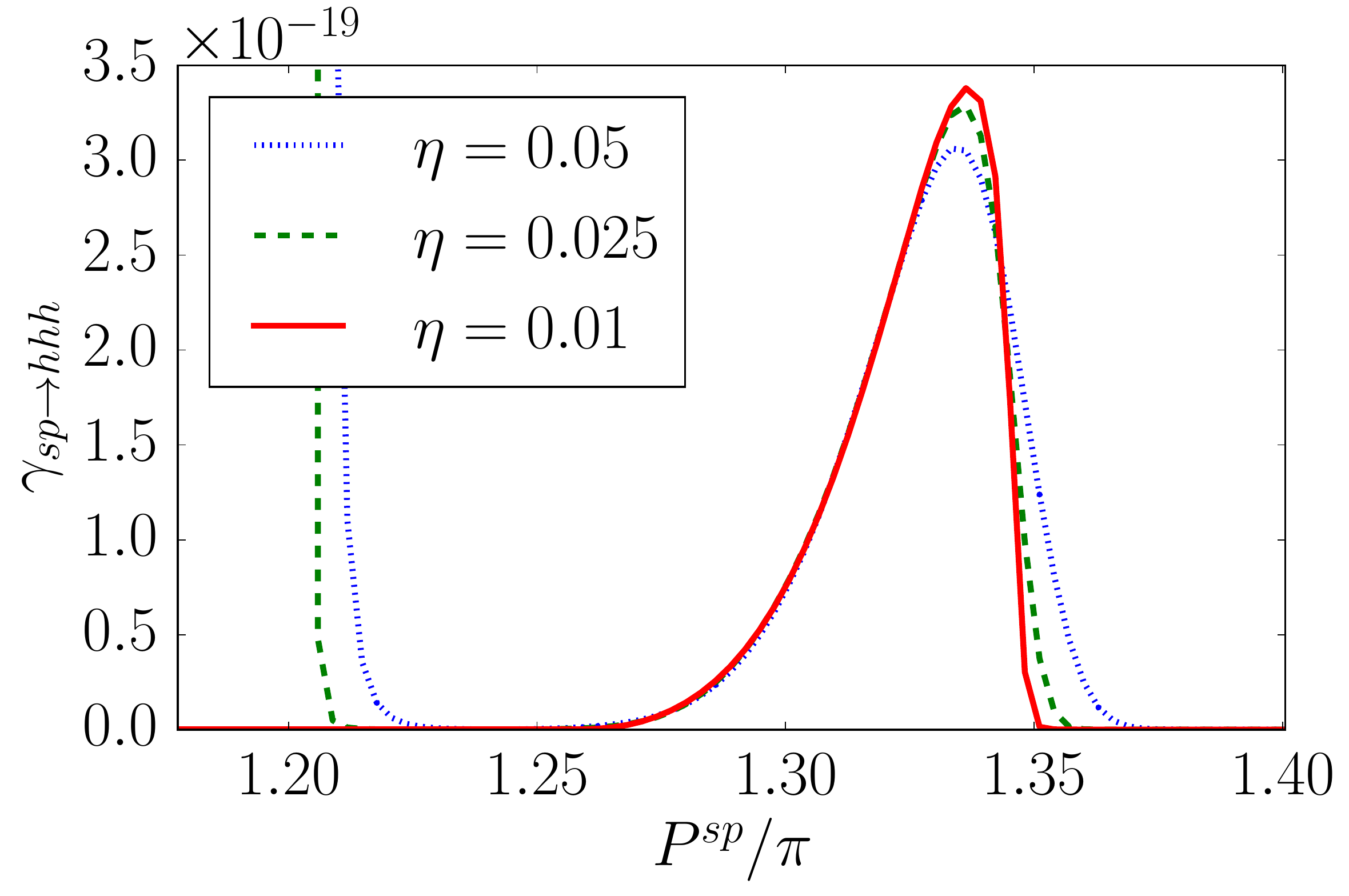} 
    \caption{}
    \label{Fig:deltadecayhhh}
  \end{subfigure} 
  \caption{Coefficients of the decay rate for magnetization $m=3/10$ and
  several values of the broadening $\eta$ for (a) pph-type processes for 
  $L=615$, (b) phh-type processes for $L=435$, (c) ppp-type processes for 
  $L=455$, (d) hhh-type processes for $L=675$.}
  \label{Fig:decay_three}
\end{figure}

We now fix $L$ and then compute \fr{decayrate} for several values of
the broadening $\eta$. The decay rates into the excitations considered
in section \ref{sec:excitations} are shown in
Figs~\ref{Fig:deltadecaypph}, \ref{Fig:deltadecayphh},
\ref{Fig:deltadecayppp} and \ref{Fig:deltadecayhhh}.

\begin{figure}[ht] 
  \begin{subfigure}{0.5\linewidth}
    \centering
    \includegraphics[width=1.0\linewidth]{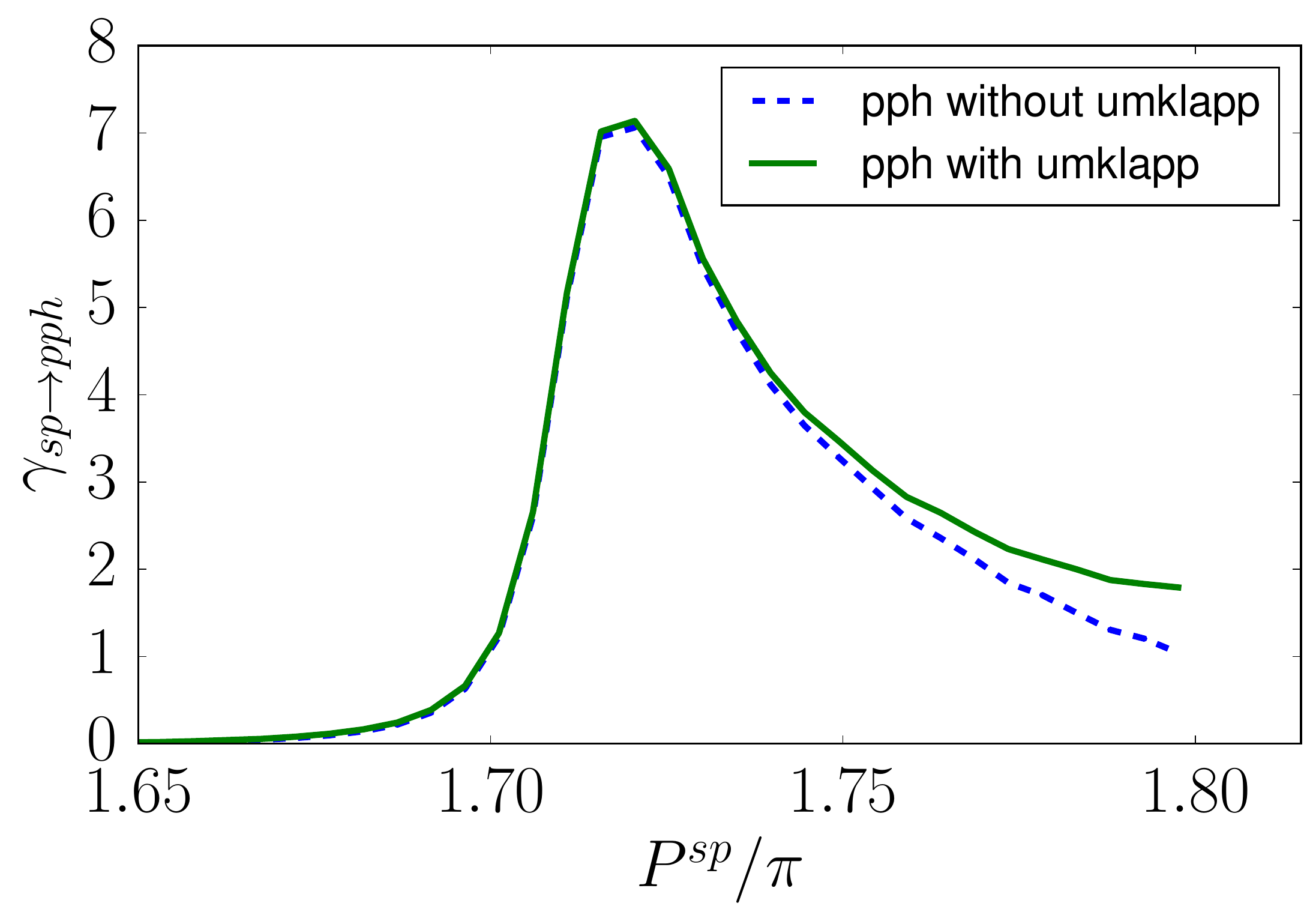} 
  \caption{}
    \label{Fig:decay_pph_umklapp}
  \end{subfigure}
  \begin{subfigure}{0.5\linewidth}
    \centering
    \includegraphics[width=1.0\linewidth]{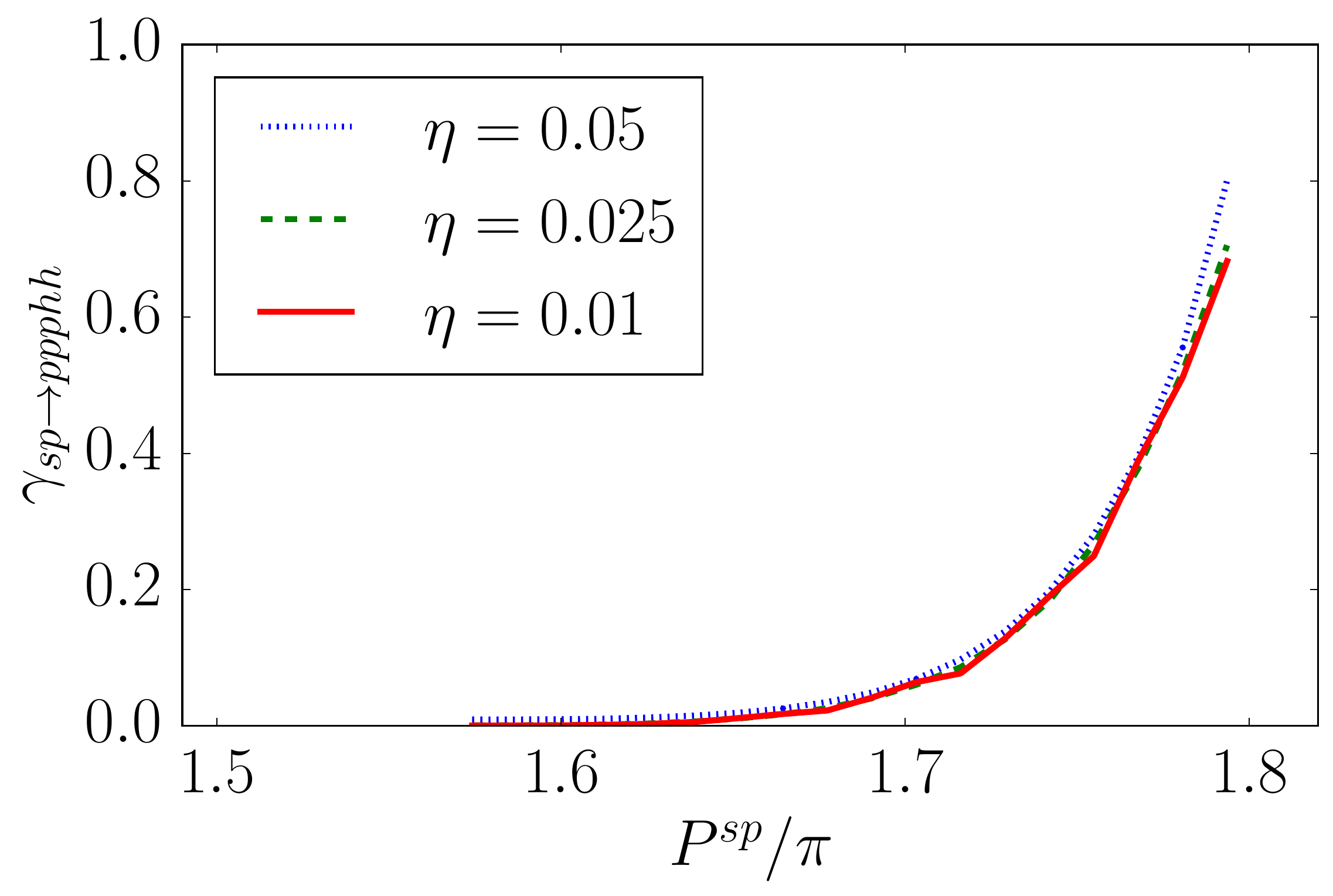} 
    \caption{}
    \label{Fig:decay_ppphh}
  \end{subfigure} 
  \caption{Coefficients of the decay rate for magnetization $m=3/10$ for
  (a) pph-type processes with and without umklapp terms for $\eta=0.002$ at 
  $L=415$ and (b) ppphh-type processes for $L=155$ and several values of the
  broadening $\eta$.}
\end{figure}

We see that the dominant decay channel for a one spinon excitation
is decay into a 3-spinon excitation of pph type. We have argued above
excitations involving higher numbers of particles should give smaller
contributions to the decay rate. In order to check this assumption we
have calculated the  decay rate into a 5-spinon excitation of type
ppphh, which we expect to provide the largest contribution among the
5-spinon excitations. The result is shown in
Fig.~\ref{Fig:decay_ppphh}. As expected the contribution is
small. Moreover, it is mostly due to umklapp-type terms in the pph-channel,
meaning particle-hole type excitations around the ``Fermi sea'' on top of the 
pph-type excitations (\emph{cf.} 
Fig.~\ref{Fig:decay_pph_umklapp}).

It is clear from Fig.~\ref{Fig:decay_three} that all other 3-spinon
decay channels can be neglected compared to the pph one.
Moreover, the decay rate coefficient is of order unity, which means
that the decay rate itself is small and proportional to the square of
the strength of the integrability breaking perturbation.

\subsection{Extrapolation to \texorpdfstring{$\eta=0$}{eta=0}}
\begin{figure}[htbp]
  \begin{center}
    \includegraphics[width=0.7\textwidth]{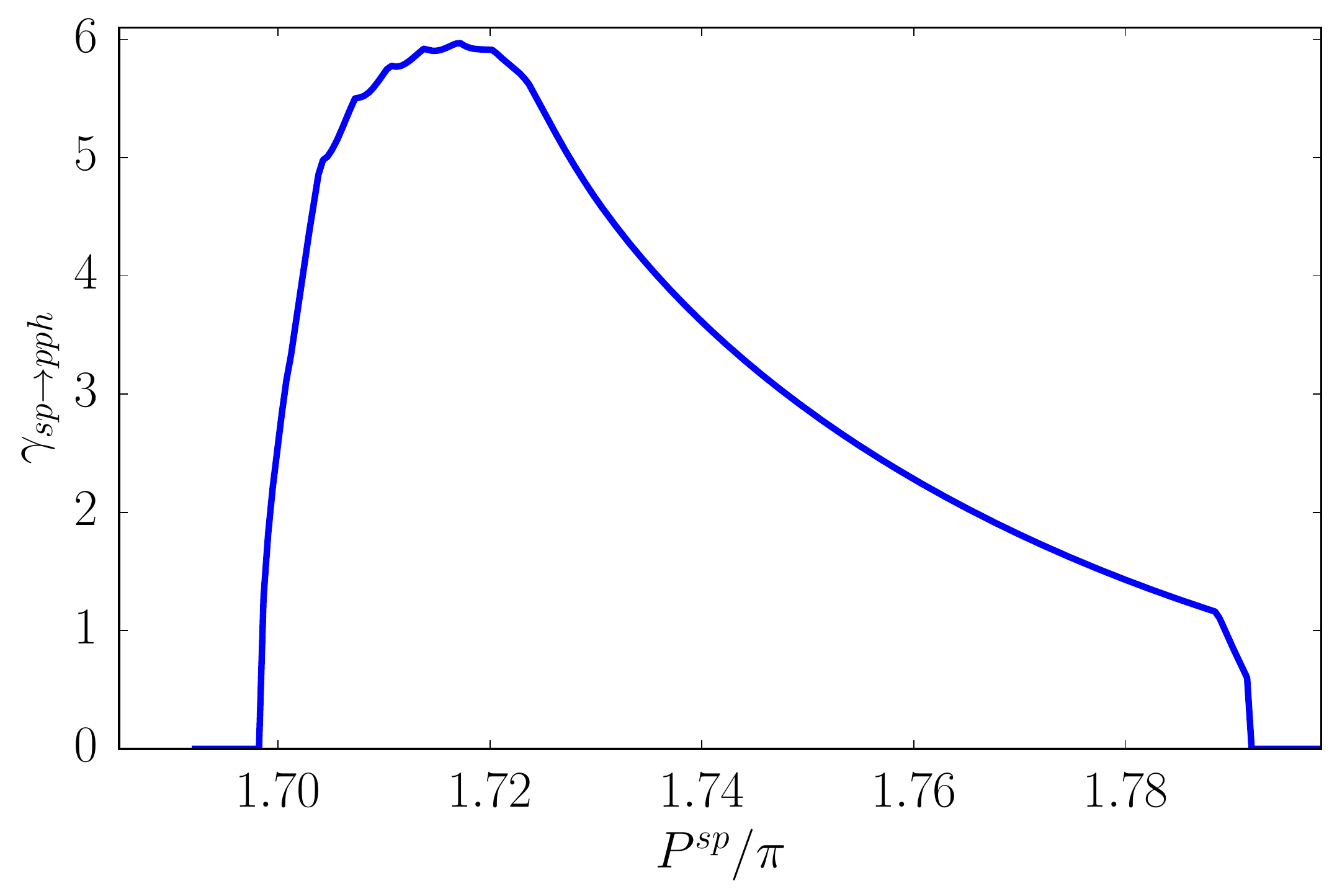}
  \end{center}
  \caption{Decay rate to the pph-channel for $\eta\to 0$ at magnetization $m=3/10$. The wiggles are due to the finite size effects of the interpolated matrix element at $L=615$.}
  \label{Fig:decay_pph_lim}
\end{figure}

The results of the previous section are for finite values of the
system size and require a small, finite regularization parameter
$\eta$. We will consider the extrapolation of these results to
the thermodynamic limit and $\eta=0$. As we have seen above, the
matrix elements of the perturbing operator scale as $L^{-1}$ up to
corrections that decay algebraically with very small exponent or
logarithmically, \emph{cf.} Fig.~\ref{Fig:logcorrections}. We expect
that in order to take the thermodynamic limit, one would have to sum
over an infinite number of particle-hole excitations at the Fermi
points, in analogy with available results for the spin-spin
correlation function
\cite{shashi1,shashi2,kitaninethermo1,kitaninethermo2,kitanineformfactor}. 
Our situation is more complicated as we need to consider excited
states with several elementary excitations at finite energies and
summing over an infinite number of particle-hole excitations on top of
these is beyond the scope of our work. However, we note that the main
source of finite-size effects in our calculation is the necessity to
have a sufficiently large value of the broadening $\eta$. This is
required in order to obtain a good approximation to density of final
states $\rho_f(E_i)$. This imposes a restriction $\eta>\eta_0(L)={\rm const}/L$.
Importantly, $\eta_0(L)$ tends to zero much faster than ${\cal
  M}^2L^2$. This allows us to extrapolate our results to $\eta=0$ as follows.
We construct a smooth interpolation function
$\mathcal{M}^{\text{cont}}_{pph}(p,z_1^p,z_2^p)$ for the matrix
element multiplied by $L$, and then turn the sums over Bethe Ansatz
(half-odd) integers into integrals using the Euler-Maclaurin sum
formula. Taking the limit $\eta\to 0$ results in 
\begin{align}
\gamma^{\rm extra}_{\rm sp\rightarrow pph}(p)&= 
\frac{1}{4\pi} \int_{D} \id{z_1^p}\int_{D}\id{z_2^p} 
\delta(\epsilon_s(z_1^p)+\epsilon_s(z_2^p)+\epsilon_s(z_1^p+z_2^p-p)-\eps(p)) \notag \\
   & \hspace{2cm}\times\abs{\mathcal{M}^{\text{cont}}_{pph}(p,z_1^p,z_2^p)}^2\ ,
\end{align}
where $D$ is the domain where the one-spinon exciation exists (the interval
$[ \pi/5,2\pi-\pi/5]$ for magnetization $m=3/10$). We stress again
that we do not claim that this integral is the exact form one would
get after summing all particle-hole pairs in the thermodynamic limit,
but that the result of such a calculation is expected to be
numerically very close to what is obtained here. One of the 
integrals can be carried out using the delta-function, which gives
\begin{align}
\gamma^{\rm extra}_{\rm sp\rightarrow pph}(p)&= 
\frac{1}{4\pi} \int_{D}
\id{z_1^p}\frac{\abs{\mathcal{M}^{\text{cont}}_{pph}(p,z_1^p,z)}^2}{\abs{\eps^\prime(z) 
+\eps^\prime(z_1^p+z-p)}}\ ,
\label{eq:finalint}
\end{align}
where $z$ is the solution of the equation 
$\epsilon_s(z_1^p)+\epsilon_s(z)+\epsilon_s(z_1^p+z-p)-\eps(p)=0$. Carrying
out the remaining integral numerically leads to the result shown in
Fig.~\ref{Fig:decay_pph_lim} 

\subsection{Density of kinematically allowed states and
  finiteness in the thermodynamic limit} 
\begin{figure}[htbp] 
  \begin{subfigure}{0.5\linewidth}
    \centering
    \includegraphics[width=1.0\linewidth]{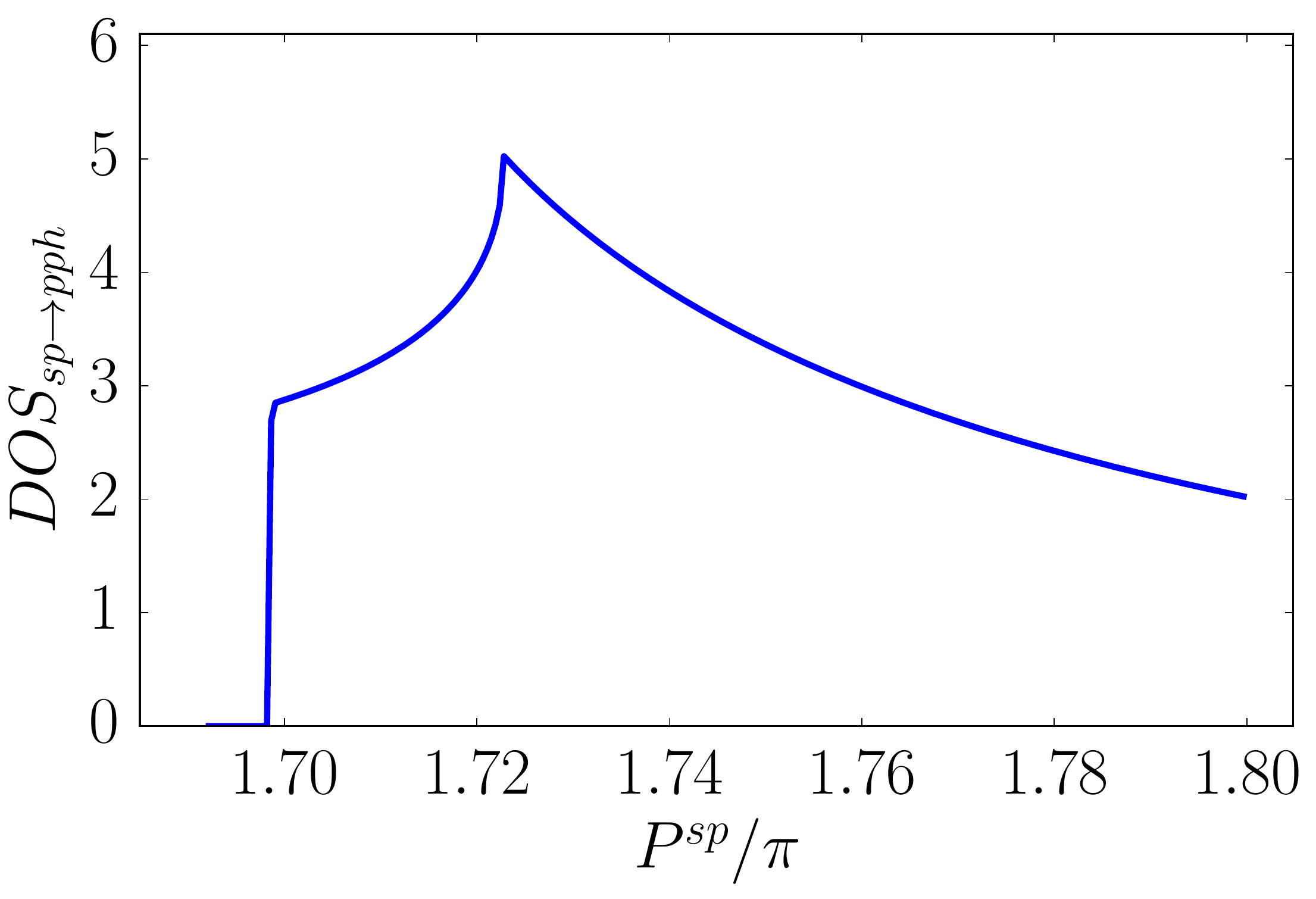} 
  \caption{}
  \end{subfigure}
  \begin{subfigure}{0.5\linewidth}
    \centering
    \includegraphics[width=1.0\linewidth]{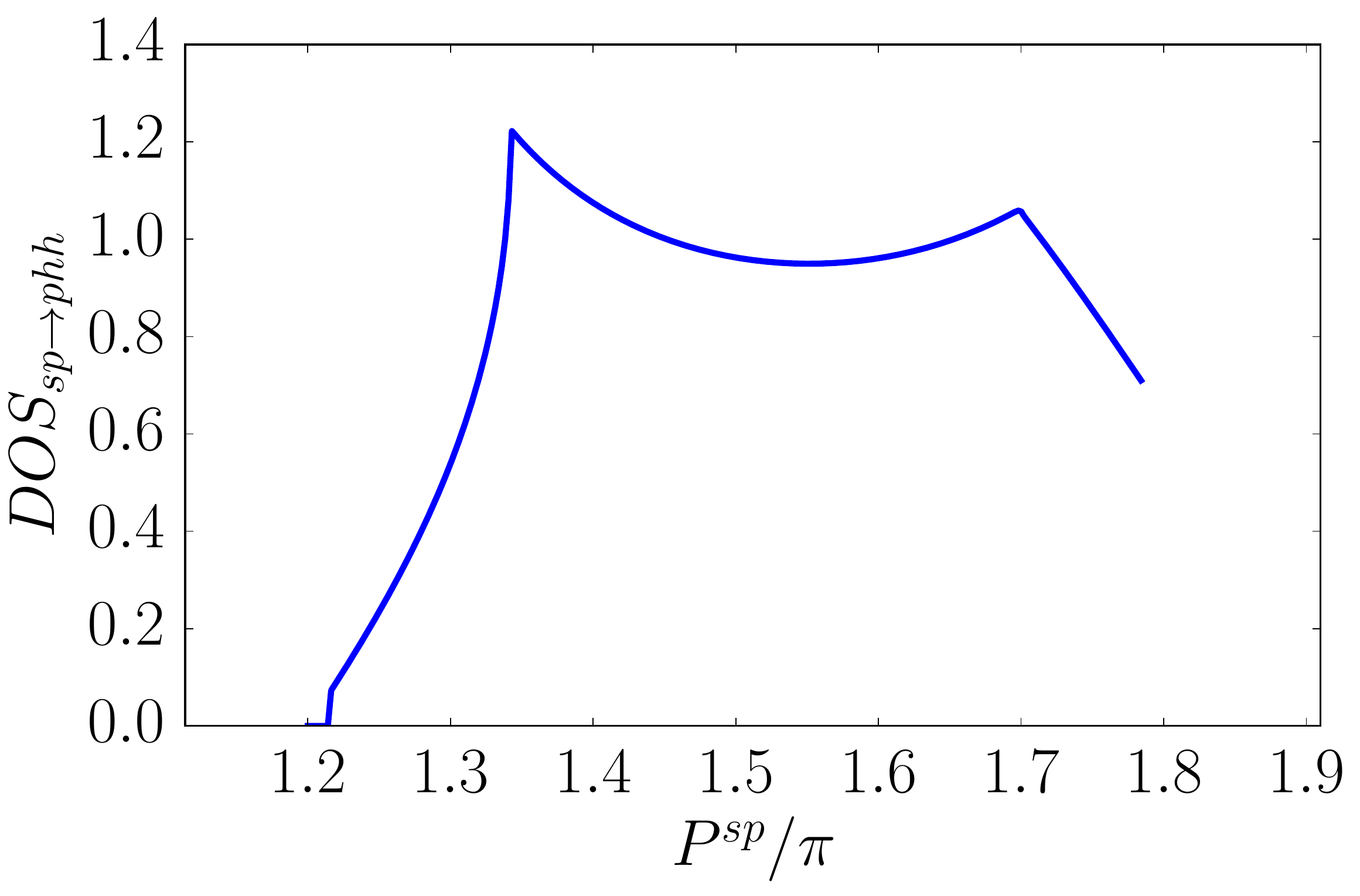} 
    \caption{}
  \end{subfigure} 
  \begin{subfigure}{0.5\linewidth}
    \centering
    \includegraphics[width=1.0\linewidth]{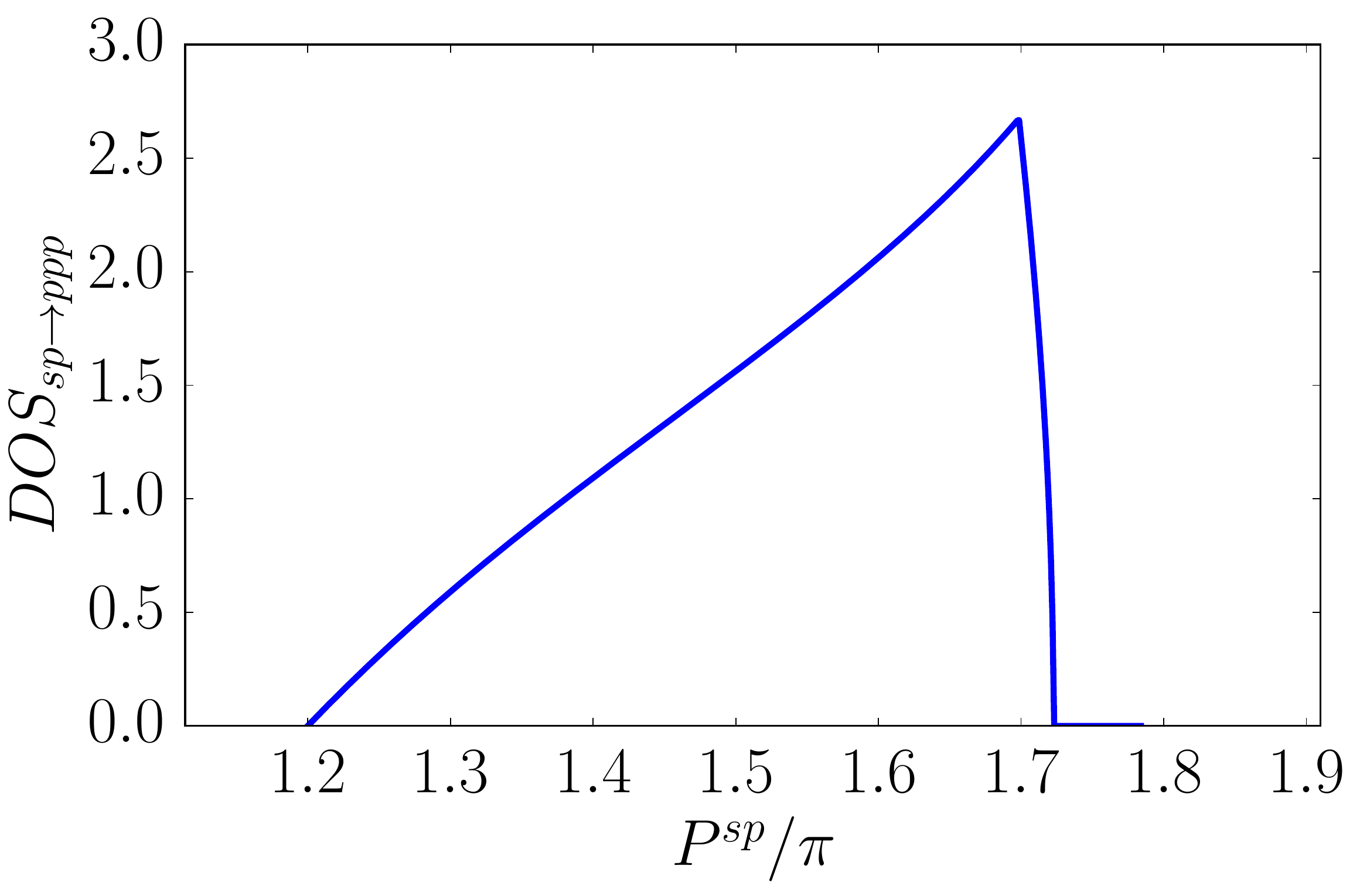} 
    \caption{}
  \end{subfigure} 
  \begin{subfigure}{0.5\linewidth}
    \centering
    \includegraphics[width=1.0\linewidth]{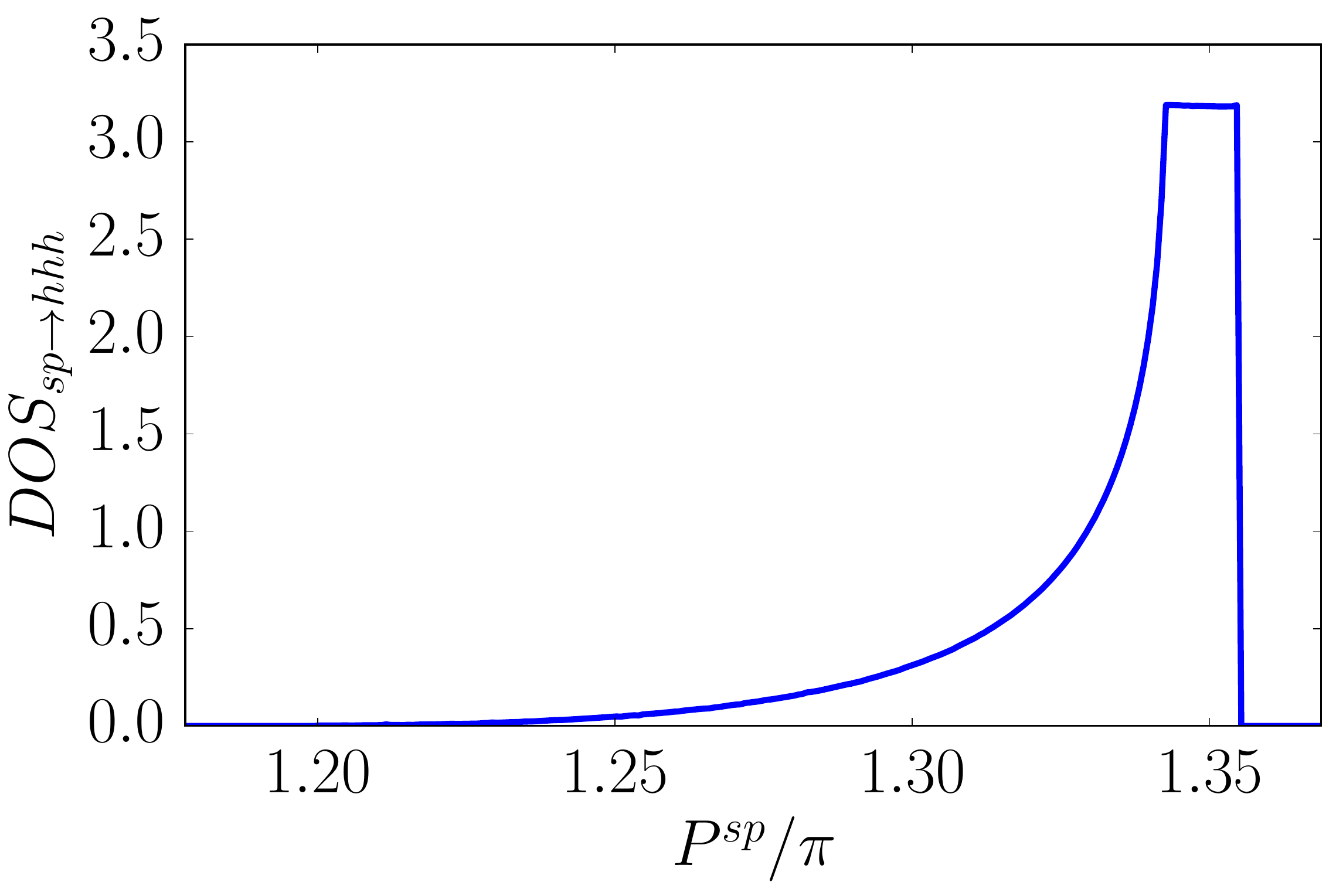} 
    \caption{}
  \end{subfigure} 

  \caption{Density of states for the (a) pph-channel (b) phh-channel, (c) 
  ppp-channel and (d) hhh-channel in the thermodynamic limit at magnetization
  $m=3/10$.}
  \label{Fig:dos_pph_lim}
\end{figure}

A simpler quantity of interest is the density of final states to
which transitions from the 1-spinon excitation are kinematically
allowed. For free fermions this density of states exhibits a van Hove
singularity that leads to logarithmic divergence at the threshold
\cite{Pereira1}. In the thermodynamic limit the pph channel density of
kinematically relevant states is given by
\begin{align}
\rho_\text{pph}(p) &= \frac{1}{4\pi} \int_{D} \id{z_1^p}\frac{1}{\abs{\eps^\prime(\tilde z)+\eps^\prime(z_1^p+z-p)}}
\end{align}
where $z$ is the same as in \fr{eq:finalint}. Analogous expressions
hold in the other channels. Results for the various possible types of
3-spinon final states are shown in Fig.~\ref{Fig:dos_pph_lim}.
We see that densities of states are finite and
do not display the kind of singularity encountered for free fermions.
\section{Conclusions}
\label{sec:summary}
We have considered decay rates of the elementary spinon excitation in
the spin-1/2 Heisenberg XXX model in a magnetic field perturbed by a weak
integrability breaking interaction $\kappa\sum_jS_j^zS^z_{j+2}$. We
have argued that the leading contribution arises from three spinon
decay and have determined the corresponding rate. The latter is found
to be small, indicating that spinons remain long-lived excitations in
the non-integrable theory. Decay of elementary string excitations can
be analyzed in an analogous fashion. This would be particularly
interesting to do in the attractive regime $\Delta<0$ of the XXZ chain
in a field, where they play an important role in the dynamics.

\section{Acknowledgments}
We are grateful to J.-S. Caux, L. Glazman and R. Pereira for helpful
discussions. This work was supported by the EPSRC under grant
EP/N01930X/1 and by the Clarendon Scholarship fund (SG).

\appendix
\section{Matrix elements and suppression for non-highest weight states} \label{Appendix:Descendants}
We want to consider the normed matrix element of $\delta H$ between a highest weight state and a non-highest weight state
\begin{align}
  \frac{\bra{\{\lambda\}}(S^+)^k\sum_j S^z_j S^z_{j+2}\ket{\{\mu\}}}{\sqrt{\matrixel{\{\lambda\}}{(S^+S^-)^k}{\{\lambda\}}}\sqrt{\braket{\{\mu\}}{\{\mu\}}}}
\end{align}
where $\bra{\{\lambda\}}$ and $\ket{\{\mu\}}$ are highest weight Bethe ansatz states. We see immediately from the 
commutation relation 
\begin{align}
  [S^+,S^z_j]=-S^+_j \label{Eq:comrelsplus}
\end{align}
and from the relation for highest weight states $S^+\ket{\{\mu\}}=0$ that for $k\geq 3$ the matrix element is exactly $0$.
Furthermore inserting the cyclic shift operator (\emph{cf.}\cite{FT}) $\exp{i\hat P}$ and using 
\begin{align}
  e^{-i\hat P} S_j^a e^{i\hat P} = S_{j+1}^a \quad a=z,+,- 
\end{align}
and the fact that the highest weight states are eigenstates of the shift operator with eigenvalue $\exp{i P_{\{\lambda\}}}$, where $P_{\{\lambda\}}$ is the momentum of the highest weight state, we obtain
\begin{align}
  \bra{\{\lambda\}}e^{-i\hat P}e^{i\hat P}&(S^+)^k\sum_j S^z_j S^z_{j+2}\ket{\{\mu\}} \notag \\ &=e^{i (P_{\{\lambda\}}-P_{\{\mu\}})} \bra{\{\lambda\}}(S^+)^k\sum_j S^z_j S^z_{j+2}\ket{\{\mu\}}
\end{align}
and therefore we see that the momenta have to conincide. 
Using the commutation relation \eqref{Eq:comrelsplus} we obtain for the normed matrix element for $k=1$ and $k=2$:
\begin{align}
  &\frac{\bra{\{\lambda\}}S^+\sum_j S^z_j S^z_{j+2}\ket{\{\mu\}}}{\sqrt{\braket{\{\mu\}}{\{\mu\}}\matrixel{\{\lambda\}}{S^+S^+S^-S^-}{\{\lambda\}}}} \notag \\
  &\qquad =\frac{1}{\sqrt{L-2(N-1)}}\frac{-\bra{\{\lambda\}}\sum_j S^+_j S^z_{j+2}+S^z_j S^+_{j+2}\ket{\{\mu\}}}{\sqrt{\braket{\{\mu\}}{\{\mu\}}\braket{\{\lambda\}}{\{\lambda\}}}} \\
  &\frac{\bra{\{\lambda\}}(S^+)^2\sum_j S^z_j S^z_{j+2}\ket{\{\mu\}}}{\sqrt{\braket{\{\mu\}}{\{\mu\}}\matrixel{\{\lambda\}}{S^+S^+S^-S^-}{\{\lambda\}}}} \notag \\
  &\qquad =\frac{1}{\sqrt{L-2(N-1)}\sqrt{L-2(N-2)}}\frac{2\bra{\{\lambda\}}\sum_j S^+_j S^+_{j+2}\ket{\{\mu\}}}{\sqrt{\braket{\{\mu\}}{\{\mu\}}\braket{\{\lambda\}}{\{\lambda\}}}}
\end{align}
We can now check numerically for solutions of the Bethe equation with same momenta using similar determinant expression as in (\emph{cf.} \ref{Appendix}) for these matrix elements, that due to the normalization factor the matrix element is suppressed for large $L$ at finite magnetic field.

\section{Calculation of the next-nearest neighbor spin operator matrix element}\label{Appendix}
We want to calculate the matrix element
\begin{align}
\bra{\{\lambda\}}\sigma_j^z \sigma^z_{j+2} \ket{\{\mu\}}
\end{align}
with $\ket{\{\mu\}}$, $\ket{\{\mu\}}$ Bethe states and $\{\mu\}$, $\{\lambda\}$ satisfying the Bethe equations \eqref{Bethesinh}. We do the calculation for all $\Delta$. To obtain the formula for $\Delta=1$ the general functions $a$ and $d$ have to be replaced for the ones mentioned above and $\gamma$ has to be set to $1$ corresponding to the rescaling of $\{\lambda\}$ with $\gamma$ and taking the limit $\gamma\to 0$.

The $\sigma$ operators are given in terms of the Bethe-Ansatz operators $A,B,C,D$, as obtained in \cite{Kitanine}:
\begin{align}
\sigma_j^z=-2\prod_{i=1}^{j-1}(A+D)(\xi_i)D(\xi_j)\prod_{k=j+1}^{L}(A+D)(\xi_k)+1
\end{align}
where $\xi_i$ is an inhomogeneity parameter, introduced at every site in the chain for technical reasons. We will set $\xi_i \to i\gamma/2$ in the end, but will keep them for the calculation. We note that now $d(\lambda)$ is defined as:
\begin{align}
d(\lambda)=\prod_{l=1}^L b(\lambda -\xi_l)
\end{align}

With this we can write the matrix element as:
\begin{align}
\bral\sigma_j^z &\sigma^z_{j+2} \ketm =\bral \sigma_j^z \ketm +\bral \sigma_{j+2}^z \ketm - \langle\{\lambda\}|\{\mu\}\rangle \notag\\
&+4\bral \prod_{i=1}^{j-1} (A+D)(\xi_i)D(\xi_j)(A+D)(\xi_{j+1})D(\xi_{j+2})\prod_{k=j+3}^{L}(A+D)(\xi_k)\ketm
\end{align}
The maxtrix elements and the overlap in the first line are known \cite{Kitanine,slavnovscalar}. However as we are interested in $\sum_j \sigma_j^z \sigma_{j+2}^z$ and as $\ket{\{\lambda\}}$ and $\ket{\{\mu\}}$ are orthogonal and eigenstates of $\sigma^z$, we only need to calculate the expression in the second line.
From the Yang-Baxter algebra one can derive the commutation relations between the operators $A,B,C,D$ and from this one gets (\cite{korepin}):
\begin{align}
A(\mu)\prod_{j=1}^N B(\lambda_j)\ket{0}&=a(\mu)\prod_{j=1}^N b^{-1}(\lambda_j-\mu)\prod_{j=1}^N B(\lambda_j)\ket{0} \notag \\ 
& -\sum_{n=1}^N a(\lambda_n) \frac{c(\lambda_n-\mu)}{b(\lambda_n-\mu)}\prod_{j\neq n}^N b^{-1}(\lambda_j-\lambda_n) B(\mu)\prod_{j\neq n}^N B(\lambda_j) \ket{0} \\
D(\mu)\prod_{j=1}^N B(\lambda_j)\ket{0}&=d(\mu)\prod_{j=1}^N b^{-1}(\mu-\lambda_j)\prod_{j=1}^N B(\lambda_j)\ket{0} \notag \\ 
& -\sum_{n=1}^N d(\lambda_n) \frac{c(\mu-\lambda_n)}{b(\mu-\lambda_n)}\prod_{j\neq n}^N b^{-1}(\lambda_n-\lambda_j) B(\mu)\prod_{j\neq n}^N B(\lambda_j) \ket{0}
\end{align}
We furthermore know that
\begin{align}
  \prod_{i=1}^j (A+D)(\xi_i)\ketl=\exp{-ij P_{\{\lambda\}} } \ketl
\end{align}
where $\ketl$ is a Bethe state and $P_{\{\lambda\}}$ is the total momentum of $\ketl$ and 
\begin{align}
  \prod_{i=1}^L (A+D)(\xi_i)\ketl=1.
\end{align}
We now need to calculate the matrix element:
\begin{align}
\mathcal{D}=\bral D(\xi_j) (A+D)(\xi_{j+1}) D(\xi_{j+2})\ketm
\end{align}
Using the commutation relations for $A,B,C,D$ (cf. \cite{korepin}) we obtain:
\begin{align}
  \mathcal{D}&=\text{\rom{1}}+\text{\rom{2}}
\end{align}
where \rom{1} is dependent on simple matrix elements where two
rapidities are replaced with inhomogeneities and \rom{2} is dependent
on a matrix element with three insertions of inhomogeneities: 
\begin{align}
\text{\rom{1}}=&\sum_{a=1}^N 
\frac{d(\mu_a)c(\xi_{j+2}-\mu_a)}{b(\xi_{j+2}-\mu_a)} 
\prod_{i\neq a}^N \frac{1}{b(\mu_a-\mu_i)}\Bigg\{ 
\frac{1}{b(\xi_{j+2}-\xi_{j+1})}\prod_{i\neq a}^N \frac{1}{b(\mu_i-\xi_{j+1})}\notag\\
&\times\ \sum_{b\neq a}^N 
\frac{d(\mu_b)c(\xi_{j}-\mu_b)}{b(\xi_{j}-\mu_b)b(\mu_b-\xi_{j+2})}
\prod_{i\neq a,b}^N \frac{1}{b(\mu_b-\mu_i)} \bral B(\xi_j)B(\xi_{j+2})\prod_{j\neq a,b}^N B(\mu_j)\ket{0} \notag \\
&+\frac{c(\xi_{j+1}-\xi_{j+2})}{b(\xi_{j+1}-\xi_{j+2})} \prod_{i\neq a}^N \frac{1}{b(\mu_i-\xi_{j+2})} \sum_{b\neq a}^N  \frac{d(\mu_b)c(\xi_{j}-\mu_b)}{b(\xi_{j}-\mu_b)}\frac{1}{b(\mu_b-\xi_{j+1})} \notag \\
&\times \prod_{i\neq a,b}^N \frac{1}{b(\mu_b-\mu_i)}\bral B(\xi_j)B(\xi_{j+1})\prod_{j\neq a,b}^N B(\mu_j)\ket{0} \Bigg\}\notag \\
 \text{\rom{2}}=&\sum_{a=1}^N 
\frac{d(\mu_a)c(\xi_{j+2}-\mu_a)}{b(\xi_{j+2}-\mu_a)} 
\prod_{i\neq a}^N \frac{1}{b(\mu_a-\mu_i)} \notag \\
&\Bigg\{\sum_{b\neq a}^N \Bigg[   \frac{c(\mu_b-\xi_{j+1})}{b(\mu_b-\xi_{j+1})} \prod_{i\neq a,b}^N
\frac{1}{b(\mu_i-\mu_b)}\frac{1}{b(\xi_{j+2}-\mu_b)}
\notag \\ &\hspace{3cm}+d(\mu_b) \frac{c(\xi_{j+1}-\mu_b)}{b(\xi_{j+1}-\mu_b)b(\mu_b-\xi_{j+2})} \prod_{i\neq a,b}^N \frac{1}{b(\mu_b-\mu_i)} \Bigg] \notag \\
& \hspace{1cm} \sum_{c\neq a,b}^N  \frac{d(\mu_c)c(\xi_{j}-\mu_c)}{b(\xi_{j}-\mu_c)b(\mu_c-\xi_{j+1})b(\mu_c-\xi_{j+2})} \prod_{i\neq a,b,c}^N \frac{1}{b(\mu_c-\mu_i)} \notag \\
& \hspace{1cm} 
\bral B(\xi_j)B(\xi_{j+1})B(\xi_{j+2})\prod_{i\neq a,b,c}^N B(\mu_i)\ket{0} \Bigg\}
\end{align}
Now we can use Slavnov's formula \cite{slavnovscalar} for the overlap of two states. One of the states has to be a Bethe state, the other state can be parametrized by an arbitrary set of rapidities. Let $\{\lambda\}$ be solutions of the Bethe equations \eqref{Bethesinh} and $\{\mu\}$ arbitrary, then one gets:
\begin{align}
\langle\{\lambda\}|\{\mu\}\rangle=\frac{\det{H(\{\lambda\},\{\mu\})}}{\prod_{j>k}\phi(\lambda_k-\lambda_j)\prod_{j<k}\phi(\mu_k-\mu_j)}
\end{align}
where $H$ is a matrix defined as
\begin{align}
H_{ab}=\frac{\phi(i\gamma)}{\phi(\lambda_a-\mu_b)}\left(a(\mu_b)\prod_{k\neq a} \phi(\lambda_k-\mu_b+i\gamma)-d(\mu_b)\prod_{k\neq a} \phi(\lambda_k-\mu_b-i\gamma)\right)
\end{align}
with $\phi(x)=x$ and $\gamma$ set to $1$ in the $\Delta=1$ scaling limit.

We will now treat \rom{1} and \rom{2} seperately. 
\subsection{Part \texorpdfstring{\rom{2}}{II}}
For the part \rom{2}, the limit of the $\xi_j,\xi_{j+1}$ and $\xi_{j+2}$ can be taken seperately for the matrix element and the prefactor. Taking the limit $\xi_i\to i\gamma/2$ for the prefactor amounts to replacing the $\xi_i$ with $i\gamma/2$. For the matrix element we obtain using Slavnov's determinant formula:
\begin{align}
  \bral B(\xi_j) &B(\xi_{j+1}) B(\xi_{j+2}) \prod_{i\neq k,m,n}B(\mu_i)\ket{0}= \notag \\ &\frac{\det{H(\{\lambda\},\{\mu_{i\neq k,m,n},\xi_j,\xi_{j+1},\xi_{j+2}\}}}{(\prod_{a>b}\phi(\lambda_a-\lambda_b)\prod_{a<b}\phi(\mu_b-\mu_a))\at{\mu_k\to\xi_j,\mu_m\to\xi_{j+1},\mu_n\to\xi_{j+2}}}
\end{align}
where 
\begin{align}
  \prod_{a<b}\phi(\mu_b-\mu_a)&\at{\mu_k\to\xi_j,\mu_m\to\xi_{j+1},\mu_n\to\xi_{j+2}} = \frac{\phi(\xi_j-\xi_{j+1})\phi(\xi_j-\xi_{j+2})\phi(\xi_{j+1}-\xi_{j+2})}{\phi(\mu_k-\mu_m)\phi(\mu_k-\mu_n)\phi(\mu_m-\mu_n)} \notag \\ & \hspace{1cm}\times\prod_{c\neq k,m,n}\frac{\phi(\xi_j-\mu_c)\phi(\xi_{j+1}-\mu_c)\phi(\xi_{j+2}-\mu_c)}{\phi(\mu_k-\mu_c)\phi(\mu_m-\mu_c)\phi(\mu_n-\mu_c)} \prod_{a<b}\phi(\mu_b-\mu_a)
\end{align}
and in the determinant we have to replace $\mu_k\to\xi_j, \mu_m\to\xi_{j+1}$ and $\mu_n\to\xi_{j+2}$. Therefore the important part when taking the limits is:
\begin{align}
  \lim_{\xi_j,\xi_{j+2}\to i\gamma/2}\frac{\det{H(\{\lambda\},\{\mu_{i\neq k,m,n},\xi_j,i\gamma/2,\xi_{j+2}\}}}{\phi(\xi_j-i\gamma/2)\phi(\xi_j-\xi_{j+2})\phi(i\gamma/2-\xi_{j+2})}
\end{align}
where the limit $\xi_{j+1}\to i\gamma/2$ is already taken. Let us now take the limit $\xi_{j+2}\to i\gamma/2$. We see that both numerator and denominator go to zero here. Therefore using the rule of l'Hospital we obtain:
\begin{align}
  &\lim_{\xi_j,\xi_{j+2}\to i\gamma/2} \frac{\det{H(\{\lambda\},\{\mu_{i\neq k,m,n},\xi_j,i\gamma/2,\xi_{j+2}\}}}{\phi(\xi_j-i\gamma/2)\phi(\xi_j-\xi_{j+2})\phi(i\gamma/2-\xi_{j+2})}= \notag \\ &\qquad\lim_{\xi_j\to i\gamma/2}  \frac{1}{\phi(\xi_j-i\gamma/2)\phi(\xi_j-i\gamma/2)}\lim_{x\to i\gamma/2}\d{}{}{x}\det{H(\{\lambda\},\{\mu_{i\neq k,m,n},\xi_j,i\gamma/2,x\})}
\end{align}
Analogous to \cite{Hagemans} we can now use a Laplace expansion of the determinant for the column that is dependent on $x$ and evaluate the derivative and limit:
\begin{align}
  &\lim_{x\to i\gamma/2}\d{}{}{x}\det{H(\{\lambda\},\{\mu_{i\neq k,m,n},\xi_j,i\gamma/2,x\})}= \notag \\ &\qquad=\lim_{x\to i\gamma/2}\d{}{}{x}\sum_{i} (-1)^{(n+i)}C_i(x) \;\text{minor}_{ni} (H(\{\lambda\},\{\mu_{i\neq k,m,n},\xi_j,i\gamma/2,x\})) \notag \\ 
  &\qquad=\lim_{x\to i\gamma/2}\sum_{i} (-1)^{(n+i)} \left(\d{}{}{x}C_i(x)\right) \;\text{minor}_{ni} (H(\{\lambda\},\{\mu_{i\neq k,m,n},\xi_j,i\gamma/2,x\}))
\end{align}
where the minor is not dependent on $x$ and $C_i(x)$ is the $i$th element of the column $n$. Therefore we get:
\begin{align}
  \lim_{x\to i\gamma/2}\d{}{}{x}&\det{H(\{\lambda\},\{\mu_{i\neq k,m,n},\xi_j,i\gamma/2,x\})}=\det{H_1(\{\lambda\},\{\mu\};m,n)\at{\mu_k\to\xi_j}}
\end{align}
with
\begin{align}
  (H_1)_{ab}(\{\lambda\},\{\mu\};m,n)&=\begin{cases} H_{ab} & b\neq m,n \\ \frac{\phi(i\gamma)}{\phi(\lambda_a-\frac{i\gamma}{2})\phi(\lambda_a+\frac{i\gamma}{2})} & b=m \\ \frac{\phi(i\gamma)\phi(2\lambda_a)}{\phi(\lambda_a-\frac{i\gamma}{2})^2\phi(\lambda_a+\frac{i\gamma}{2})^2} & b=n \end{cases}
\end{align}
Repeating this step for the limit $\xi_j\to i\gamma/2$ using the rule of l'Hospital twice we finally obtain:
\begin{align}
  M_{ijk}^{(2)} &\equiv \lim_{\xi_j,\xi_{j+1},\xi_{j+2}\to i\gamma/2}\bral B(\xi_j) B(\xi_{j+1}) B(\xi_{j+2}) \prod_{i\neq k,m,n}B(\mu_i)\ket{0} \notag \\
  &=\frac{1}{2}\prod_{i} \phi\left(\lambda_i+\frac{i\gamma}{2}\right)^3 \phi(\mu_m-\mu_n)\phi(\mu_k-\mu_m)\phi(\mu_k-\mu_n) \notag \\
  &\quad\times\prod_{c\neq k,m,n}\frac{\phi(\mu_c-\mu_k)\phi(\mu_c-\mu_m)\phi(\mu_c-\mu_n)}{\phi(\mu_c-\frac{i\gamma}{2})^3} \notag \\ 
  &\quad\times\frac{\det{H_2}}{\prod_{a>b}\phi(\lambda_b-\lambda_a)\prod_{a<b}\phi(\mu_b-\mu_a)}
\end{align}
where 
\begin{align}
  (H_2)_{ab}(\{\lambda\},\{\mu\};k,m,n)&=\begin{cases} H_{ab} & b\neq m,n,k \\ \frac{\phi(i\gamma)}{\phi(\lambda_a-\frac{i\gamma}{2})\phi(\lambda_a+\frac{i\gamma}{2})} & b=m \\ \frac{\phi(i\gamma)\phi(2\lambda_a)}{\phi(\lambda_a-\frac{i\gamma}{2})^2\phi(\lambda_a+\frac{i\gamma}{2})^2} & b=n  \\ \pd{2}{}{x} \left( \frac{\phi(i\gamma)}{\phi(\lambda_a-x)\phi(\lambda_a-x+i\gamma)}\right)\at{x=\frac{i\gamma}{2}} & b=k\end{cases}
\end{align}
With this we obtain after some algebra:
\begin{align}
  \text{\rom{2}}&=-\frac{1}{2} \frac{\prod_k\phi(\lambda_k+i\gamma/2)^3\phi(\mu_k-i\gamma/2)^{-3}}{\prod_{a>b}\phi(\lambda_b-\lambda_a)\prod_{a<b}\phi(\mu_b-\mu_a)}\sum_{n=1}^N d(\mu_n) \phi(\mu_n-i\gamma/2)\prod_i \phi(\mu_i-\mu_n-i\gamma) \notag \\ 
  &\quad \sum_{m\neq n} \prod_i \phi(\mu_i-\mu_m+i\gamma)\phi(\mu_m-i\gamma/2)\left[ \frac{\phi(\mu_m-3i\gamma/2)}{\phi(\mu_m-\mu_n-i\gamma)}-\frac{\phi(\mu_m+i\gamma/2)}{\phi(\mu_m-\mu_m+i\gamma)}\right] \notag \\ 
  &\quad \sum_{k\neq m,n} d(\mu_k)\prod_i \phi(\mu_i-\mu_k-i\gamma) \frac{\phi(\mu_k+i\gamma/2)^2}{\phi(\mu_m-\mu_k-i\gamma)\phi(\mu_n-\mu_k-i\gamma)} \notag \\ & \hspace{3cm} \times\det{H_2(\{\lambda\},\{\mu\};k,m,n)}
\end{align}
Using a Lemma from Laplace's determinant formula (\cite{Klauser}) we finally obtain:
\begin{align}
  \text{\rom{2}}&=-\frac{1}{2} \frac{\prod_k\phi(\lambda_k+i\gamma/2)^3\phi(\mu_k-i\gamma/2)^{-3}}{\prod_{a>b}\phi(\lambda_b-\lambda_a)\prod_{a<b}\phi(\mu_b-\mu_a)} \notag\\ &\qquad\times\sum_{n=1}^N \sum_{m\neq n} A_{nm} \left[\det{G_{nm}+B_{nm}}-\det{G_{nm}}\right]
\end{align}
with 
\begin{align}
  (G_{nm})_{ab}&=\begin{cases} H_{ab} & b\neq m,n \\ \frac{\phi(i\gamma)\phi(2\lambda_a)}{\phi(\lambda_a-\frac{i\gamma}{2})^2\phi(\lambda_a+\frac{i\gamma}{2})^2} & b=m  \\ \pd{2}{}{x} \left( \frac{\phi(i\gamma)}{\phi(\lambda_a-x)\phi(\lambda_a-x+i\gamma)}\right)\at{x=\frac{i\gamma}{2}} & b=n \end{cases} \\ 
    (B_{nm})_{ab}&=(1-\delta_{bm})(1-\delta_{bn})d(\mu_b) \notag \\ & \hspace{1cm}\prod_{i\neq m,n}\phi(\mu_i-\mu_b-i\gamma) \phi(\mu_b+i\gamma/2)^2 \frac{\phi(i\gamma)}{\phi(\lambda_a-i\gamma/2)\phi(\lambda_a+i\gamma/2)}
\end{align}
\subsection{Part \texorpdfstring{\rom{1}}{I}}
First we can take the limit $\xi_j,\xi_{j+2}\to i\gamma/2$. This again just amounts to replacing the $\xi_j$ and $\xi_{j+2}$ with $i\gamma/2$ in the prefactors and doing a similar analysis for the matrix element depending on both $\xi_j$ and $\xi_{j+2}$ as for the part \rom{2}. However performing the limit $\xi_{j+1}\to i\gamma/2$ is a bit more involved, as the limit can not be taken independently for prefactor and matrix elements. This is due to terms $\propto \frac{1}{\xi_{j+1}-i\gamma/2}$ appearing in the prefactors. Doing a consistent series expansion in $\xi_{j+1}-i\gamma/2$ again utilizing the Laplace determinant expansion and then taking the limit $\xi_{j+1}\to i\gamma/2$ we obtain after some calculation:
\begin{align}
  \text{\rom{1}}&=-\phi(i\gamma)\frac{\prod_k\phi(\lambda_k+i\gamma/2)^2\phi(\mu_k+i\gamma/2)\phi(\mu_k-i\gamma/2)^{-3}}{\prod_{a>b}\phi(\lambda_b-\lambda_a)\prod_{b>a}\phi(\mu_b-\mu_a)} \notag \\ &\left\{ \sum_{n=1}^N d(\mu_n)\prod_i \phi(\mu_i-\mu_n-i\gamma) \frac{\phi(\mu_n-i\gamma/2)^2}{\phi(\mu_n+i\gamma/2)} \right.\notag \\
  & \times \left[ \sum_{m\neq n} d(\mu_m) \prod_{i\neq n} \phi(\mu_i-\mu_m-i\gamma)\phi(\mu_m+i\gamma/2) \right. \notag \\ &\left( \frac{\phi^\prime(i\gamma)}{\phi(i\gamma)}-\sum_{d\neq n} \frac{\phi(i\gamma)}{\phi(\mu_d-i\gamma/2)\phi(\mu_d+i\gamma/2)}+\sum_{d\neq m,n} \frac{\phi^\prime(\mu_d-i\gamma/2)}{\phi(\mu_d-i\gamma/2)}\right.\notag \\ &-\sum_b \frac{\phi^\prime(\lambda_b+i\gamma/2)}{\phi(\lambda_b+i\gamma/2)}+\left.\frac{\phi(i\gamma)}{\phi(\mu_m+i\gamma/2)\phi(\mu_m-i\gamma/2)}\right) \det{H_3(\{\lambda\},\{\mu\};m,n)}+ \notag \\ 
  & \left. \left. \sum_{m\neq n} d(\mu_m)\phi(\mu_m+i\gamma/2)  \prod_{i\neq n} \phi(\mu_i-\mu_m-i\gamma)\frac{1}{2} \det{H_4(\{\lambda\},\{\mu\};m,n)}\right]\right\}
\end{align}
with 
\begin{align}
  (H_3)_{ab}&=\begin{cases} H_{ab} & b\neq m,n \\ \frac{\phi(i\gamma)}{\phi(\lambda_a+i\gamma/2)\phi(\lambda_a-i\gamma/2)}& b=m \\ \frac{\phi(i\gamma)\phi(2\lambda_a)}{\phi(\lambda_a-\frac{i\gamma}{2})^2\phi(\lambda_a+\frac{i\gamma}{2})^2} & b=n  \end{cases} \\  
  (H_{4})_{ab}&=\begin{cases} H_{ab} & b\neq m,n \\ \frac{\phi(i\gamma)}{\phi(\lambda_a+i\gamma/2)\phi(\lambda_a-i\gamma/2)}& b=m  \\ \pd{2}{}{x} \left( \frac{\phi(i\gamma)}{\phi(\lambda_a-x)\phi(\lambda_a-x+i\gamma)}\right)\at{x=\frac{i\gamma}{2}} & b=n \end{cases} 
\end{align}
using the Lemma from Laplace's determinant formula again, we finally obtain:
\begin{align}
 \text{\rom{1}}&=-\phi(i\gamma) \frac{\prod_k\phi(\lambda_k+i\gamma/2)^2\phi(\mu_k+i\gamma/2)\phi(\mu_k-i\gamma/2)^{-3}}{\prod_{a>b}\phi(\lambda_b-\lambda_a)\prod_{b>a}\phi(\mu_b-\mu_a)} \notag \\
 & \quad\times\sum_n A_n \Big(\det{G_n^{(1)}+B_n^{(1)}}+\det{G_n^{(2)}+B_n^{(2)}} \notag \\
 & \qquad \qquad\qquad -\det{G^{(1)}}-\det{G^{(2)}}\Big)
\end{align}
where
\begin{align}
  (G^{(1)}_n)_{ab} &= \begin{cases} H_{ab} & b\neq n \\ \frac{\phi(i\gamma)\phi(2\lambda_a)}{\phi(\lambda_a-\frac{i\gamma}{2})^2\phi(\lambda_a+\frac{i\gamma}{2})^2} & b=n \end{cases} \\
  (G^{(2)}_n)_{ab} &= \begin{cases} H_{ab} & b\neq n \\ \pd{2}{}{x} \left( \frac{\phi(i\gamma)}{\phi(\lambda_a-x)\phi(\lambda_a-x+i\gamma)}\right)\at{x=\frac{i\gamma}{2}} & b=n \end{cases} \\
  (B^{(1)}_n)_{ab} &= (1-\delta_{bn}) d(\mu_b) \prod_{i\neq n} \phi(\mu_i-\mu_b+i\gamma)\phi(\mu_b+i\gamma/2) \notag \\ &\left( \frac{\phi^\prime(i\gamma)}{\phi(i\gamma)}-\sum_{d\neq n} \frac{\phi(i\gamma)}{\phi(\mu_d-i\gamma/2)\phi(\mu_d+i\gamma/2)}+\sum_{d\neq m,n} \frac{\phi^\prime(\mu_d-i\gamma/2)}{\phi(\mu_d-i\gamma/2)} \right.\notag \\ &\left. -\sum_b \frac{\phi^\prime(\lambda_b+i\gamma/2)}{\phi(\lambda_b+i\gamma/2)}+\frac{\phi(i\gamma)}{\phi(\mu_m+i\gamma/2)\phi(\mu_m-i\gamma/2)}\right) \\
  (B^{(2)}_n)_{ab} &= (1-\delta_{bn}) \frac{1}{2} d_{\mu_b}  \notag \\ & \hspace{1cm} \times\prod_{i\neq n} \phi(\mu_i-\mu_b-i\gamma)\phi(\mu_b+i\gamma) \frac{\phi(i\gamma)}{\phi(\lambda_a+i\gamma/2)\phi(\lambda_a-i\gamma/2)}
\end{align}
\subsection{Total matrix element}
We can now put together the total matrix element. We obtain
\begin{align}
  \sum_j\bral S_j^z S_{j+2}^z \ketm &=e^{iP_{\{\lambda\}}+2iP_{\{\mu\}}}\sum_j e^{-ij (P_{\{\lambda\}}-P_{\{\mu\}})} \mathcal{D} \\
  &=L e^{iP_{\{\lambda\}}+2iP_{\{\mu\}}} \delta_{P_{\{\lambda\}},P_{\{\mu\}}} \mathcal{D}
\end{align}


\end{document}